\documentclass{acmsmall}
\usepackage[ruled,vlined]{algorithm2e}
\title{Formal and Computational Properties of the \\
Confidence Boost of Association Rules%
}
\author{JOS\'E L. BALC\'AZAR
\affil{Universidad de Cantabria}
}
\newcommand{\D}{\mathcal{D}}
\newcommand{\U}{\mathcal{U}}
\newcommand{\R}{\mathcal{R}}
\newcommand{\B}{\mathcal{B}}
\newcommand{\Bst}[1]{{\B}^{\star}_{#1}}
\newcommand{\cl}[1]{\overline{#1}}
\newcommand{\implies}{\to}
\newcommand{\st}{\bigm|}
\def\qed{\nobreak\kern1ex\vrule width4pt height4pt depth0pt}
\newcommand\ds[1]{\mbox{\sc#1}}

\begin{document}

\terms{Algorithms, Theory, Human factors}
\acmformat{}
\begin{bottomstuff}
Address at the time of submission:
Departmento de Matem\'aticas, Estad\'\i{}stica y Computaci\'on,
Av Los Castros s/n, Santander 39005, Spain 
({\tt joseluis.balcazar}@{\tt unican.es}).
 This work has been partially supported by project TIN2007-66523
 (FORMALISM)
 of Programa Nacional de Investigaci\'on, Ministerio de Ciencia
 e Innovaci\'on (MICINN), Spain, and by the Pascal-2 Network of the 
 European Union.
\end{bottomstuff}
\begin{abstract} 
Some existing notions of redundancy 
among association rules allow for a logical-style 
characterization and lead to irredundant bases 
of absolutely minimum size. 
One can push the intuition of redundancy further 
and 
find an intuitive notion of interest of 
an association rule, in terms of its ``novelty'' with 
respect to other rules. 
Namely: an irredundant rule is so 
because its confidence is higher than what the rest 
of the rules would suggest; then, one can ask: 
how much higher? 

We propose to measure such a sort of ``novelty''
through the confidence boost of a rule, which
encompasses two previous similar notions 
(confidence width and rule blocking, of which
the latter is closely related to the earlier measure
``improvement''). 
Acting as a complement to confidence
and support, the confidence boost 
helps to obtain small and crisp sets of
mined association rules, and solves the well-known
problem that, in certain cases, rules of negative
correlation may pass the confidence bound.
We analyze the properties of two versions of 
the notion of confidence boost, 
one of them a natural generalization of the other.
We develop efficient algorithmics to filter rules 
according to their confidence boost, compare 
the concept to 
some similar notions in the bibliography, and describe 
the results of some experimentation employing the new
notions on standard benchmark datasets. 
We 
describe an open-source 
association mining 
tool that embodies one of our
variants of confidence boost 
in such a way 
that the data mining process
does not require the user to select any value for
any parameter.
%
%
%
\end{abstract}

\maketitle

\section{Introduction}

As the now well-known task of association rule mining
was defined, the problems faced were twofold. 
First, the quantity of candidate itemsets for 
antecedent $X$ and consequent $Y$ of 
association rules $X\to Y$ grows 
exponentially with the often already large universe of items. 
The introduction of a \emph{support threshold} parameter was a key advance that 
allowed for the design of efficient frequent set miners and for 
the computation of association rules in large datasets: there,
exploration is limited to those itemsets that appear ``often enough"
as subsets of the transactions, that is, their relative
frequency exceeds a certain ratio of the transactions;
see \cite{AMSTV} and the references there.
Then, the second problem is that, often, the set of 
rules provided as output is too large, specially if we 
consider that its purpose is to be read, and understood, by a human. 
We consider that this problem warrants further research,
and we attempt at providing here yet one more approach to it.

These two difficulties are of very different sorts. The
exponential growth of candidates is essentially a combinatorial,
almost technological problem, and all the existing solutions are 
based on the acceptance that, as not all the billions of 
candidates can be considered within reasonable running
times, we make do with those that obey the support constraint.
However, this solution puts unto the shoulders of the user 
the heavy responsibility of choosing the support threshold, 
with little or no guidance about how to do it.

On the other hand, it is no problem for our current computing
equipments to extract association rules from frequent sets.
The proposal in \cite{AMSTV} (and already in the early \cite{Lux}
where, however, the support bound proposal does not appear) 
is to impose upon association rules $X\to Y$ a confidence 
constraint, that is, a threshold on the conditional probability 
of $Y$ conditioned to $X$.

Indeed,
association rule mining, in essence, amounts to enumerating
all the rules that are not disproved by the data.
As there are exponentially
growing quantities of potential associations, even relatively
large datasets are unable to disprove most of them.
Therefore,
in the standard ``support and confidence'' framework, it is well-known,
and easy to check using any of the public datasets and free association
miners available on the web, that whereas high, demanding thresholds for 
these parameters generally yield few somewhat obvious rules, 
softening them, as much as the algorithmics (and the user patience)
would allow, leads to large amounts of rules, with many of
them looking very much like each other; often, they are
not a user-friendly enough result of a data mining process,
due to the presence of these intuitive redundancies.

As a preliminary filter, there are several essentially logical
definitions of redundancy, patterned after similar intuitions
in Propositional or First-Order Logic. This leads to minimum-size
bases, such as the Representative (or Essential) Rules 
\cite{AgYu,KryszPAKDD}
for 
plain redundancy or the basis $\Bst{\gamma}$ 
\cite{Bal10b} for closure-based redundancy,
at confidence threshold $\gamma$, 
that 
spares the computation of minimal generators needed by the Representative
Rules, but needs to be complemented with a basis for full implications.
All these questions are thoroughly
surveyed in \cite{Bal10b}.
But even taking
redundancies into account, the results are, in many cases,
unsatisfactory; therefore,
many alternative quality measures exist for
association rules, essentially due to the facts that, 
first, the confidence of a rule $X\to Y$ can be high even 
in cases where the actual correlation between $X$ and $Y$ 
is negative, and, second, it is often extremely difficult 
to settle for thresholds where interesting rules are kept 
but the total amount of rules can be handled; see 
\cite{GH,LencaEtAl,TKS} and their references for 
information about the rich research area opened up by these 
difficulties. We note
that,
from the point of view of the user,
the usage of alternative implicational measures leads to an
even worse situation, as (s)he has 
to choose again both the measures to apply and their
corresponding thresholds.
The literature on this topic is huge and cannot be
reviewed here; a discussion
of the relationships of 
our contributions 
with the 
most relevant ones among 
the published proposals is
deferred to Subsections \ref{ss:relwork}~and~\ref{ss:mmr}.

Our development is based on the simple consideration 
that
rules can be
evaluated for ``novelty'', 
by comparison with the rest of the rules mined.
Actually, the outcome of 
every Data Mining project 
is expected to offer some degree of novelty. 
If one ends up identifying only 
facts whose validity is obvious,
these
would not be really useful. 
However, to 
formally study the novelty of Data Mining results is far from being
a trivial task.
Indeed, novelty is, in an intuitive sense, a relative notion: 
it refers to facts that are, somehow, unexpected; hence, 
some ``low expectation'' reason must exist,
and must be due to alternative facts or prediction mechanisms. 
That is: a piece of information is novel or is not, always 
with respect to a given context of previously known facts;
definitions of novelty must take into account, then, 
some form of previously available knowledge, a notion hard to 
formalize (Subsection~\ref{ss:relwork} describes some approaches,
but see~e.g.~\cite{PadTu2000} and the references therein).

However, as one very partial and probably insufficient, but necessary
action, we claim that, as a minimum, each rule should be
evaluated for novelty by comparison with 
the rest of the rules mined, 
treated as ``alternative'' mechanism~\cite{Bal09}.
%
One can attempt at measuring
to what extent the confidence
of the rule is substantially higher than that of related rules that
would, intuitively, explain the same facts. 
In
the same reference,
the {\em confidence width} is proposed as a measure of a relative 
form of objective novelty or surprisingness of each individual rule 
with respect to other rules that hold in the same dataset. 
As some intuitive redundancies are not covered by that measure, 
the same paper 
proposes also to allow 
some rules to {\em block} other rules in case the blocked rule does not 
bring in enough novelty with respect to the blocker. (We give below
the precise definitions of these notions.)
Essentially, these 
proposals measure novelty through the extent to which the confidence 
value is ``robust'', taken relative to the confidences of related rules, 
as opposed to the absolute consideration of the single rule at hand. 

To give a hint of the sort of process we are discussing, assume
a rule, of confidence say 75\%, is found in a census-like dataset,
stating that young people earn lesser salaries; in the presence of 
such a rule, a more complex one stating that young, unmarried
people earn lesser salaries could be novel, but only if its confidence
turns out to be substantially higher than 75\%, maybe 90\%. Otherwise,
it would not be novel, the simpler rule should be preferred, and even
the complex rule discarded (or \emph{blocked}), 
all depending on thresholds on confidence
and on some other parameter such as improvement \cite{BAG},
blocking factor, 
or confidence boost (to be introduced here). 
Further discussion will be provided along
the body of the paper.

It was empirically demonstrated in \cite{Bal09} that better results
were obtained using both a confidence width threshold and a blocking
threshold, than using a single one of these filters (or~none). However,
no really fast way of testing a rule for blockings was provided.
Thus, our contribution here is a new attempt at formalizing the notion
of novelty, the {\em confidence boost}, 
similar in its syntactic definition to confidence width, but different 
in its semantics, which is more restrictive; its main feature is that 
it encompasses at once both the bound on the confidence width and the 
ability to detect that a rule would be blocked, so that the confidence
boost bound embodies both of the bounds proposed in~\cite{Bal09}, yet
it is computable with reasonable efficiency.
Confidence boost comes in two flavors: a ``plain'' one, that
we develop in Section~\ref{s:cb}, 
and a more general
variant that takes into account the closure space implicit in the data,
developed in Section~\ref{s:cbcb}.

Three short extended abstracts of three, six, and seven pages 
respectively have announced
results from this paper in scientific meetings; reference~\cite{Bal10a}
contains the definition of confidence boost, fragments of Section~\ref{s:prel}
(where we also review a small number of necessary facts from \cite{Bal10b}),
part of Section~\ref{s:cb} (the definition of confidence boost),
and the
algorithm in Subsection~\ref{ss:algo} (but not its correctness proof).
Reference~\cite{Bal10c} contains the definition of closure-based 
confidence boost and part of the materials in Section~\ref{s:cbcb},
again including the main algorithm but not its correctness proof, 
as well as materials from Subsection~\ref{ss:subjeval}. 
The tool {\sl yacaree} which embodies closure-based confidence
boost into a parameter-free association miner 
(Section~\ref{s:yacaree}) was advertised
at \cite{Bal11a} (demo~track). The rest
of Sections \ref{s:cb}, \ref{s:cbcb},
and~\ref{s:emval}, as well as the discussions 
in Section~\ref{s:discussion}, are unpublished.


\section{Preliminaries}
\label{s:prel}

A given set of available items $\U$ 
is assumed; its subsets are
called itemsets.
We will denote itemsets by capital letters from the end of the
alphabet, and use juxtaposition to denote union, as in $XY$. 
The inclusion sign as in $X\subset Y$ denotes proper subset,
whereas improper inclusion is denoted $X\subseteq Y$.
For a given dataset~$\D$, consisting of~$n$ transactions,
each of which is an itemset labeled with a unique transaction
identifier, we can count the \emph{support} $s_{\D}(X)$ of an itemset $X$,
which is the cardinality of the set of transactions that
contain~$X$. An alternative rendering of support is its
normalized version, the relative frequency or
empirical probability $s_{\D}(X)/n$; we will work with the
unnormalized quantity.

Association miners explore datasets in search 
of valid expressions of the form $X\to Y$, where $X$ and $Y$
stand for itemsets.
Intuitively, an association rule $X\to Y$ means that, in the
given dataset, the transactions that contain $X$ 
``tend to contain''~$Y$ as well. 
The \emph{confidence} of a 
rule $X\to Y$ 
is $c_{\D}(X\to Y) = s_{\D}(XY)/s_{\D}(X)$, 
akin to an empirical approximation to a
conditional probability.
It is important to observe that
the precise definition of association rules
depends on the formalization chosen for the informal 
expression ``tend to'', as only then these syntactical
expressions become endowed with a concrete semantics and associated 
specific properties.
For instance, if we define the meaning of $X\to Y$ through
confidence, then rules $X\to Y$ and $X\to XY$ are equivalent,
whereas, if we use lift (defined below), then they may not
be equivalent.

Confidence is a very natural notion to prune and rank
the output of an association rule mining algorithm,
but we must point out that,
due to some objections that we review in Subsection~\ref{ss:block},
there exist other proposals of notions to replace
confidence. 
When confidence is~1, the maximum value, we
say that $X\to Y$ is an \emph{implication}: 
every transaction containing $X$ contains as well $Y$.
Sometimes we use the term \emph{partial rule} for
an association rule of confidence less than~1.
The \emph{support} of a rule $X\to Y$ 
is $s_{\D}(X\to Y) = s_{\D}(XY)$. 
When the dataset is clear from the context, we will omit
the subscript $\D$ from both support and confidence.
We do allow $X=\emptyset$ as antecedent of association rules:
then the confidence coincides with the normalized support,
$c(\emptyset\to Y) = s(Y)/s(\emptyset) = s(Y)/n$.
Allowing $Y=\emptyset$ as consequent as well is possible
but not very useful, as this case leads only to trivial 
rules equivalent to reflexivity statements; therefore
we assume that such rules are omitted from all our
sets of rules.
In the proposal of \cite{AMSTV},
association rules are restricted to $|Y|=1$. This 
allows for faster algorithmics, as rules are directly
obtained from each frequent set. In fact, whereas
confidence~1 implications, say, $A\to B$ and $A\to C$ 
jointly are indeed
equivalent to $A\to BC$, for confidence less than 1
they are not. $A\to BC$ says that $B$ and $C$ appear
\emph{jointly} often with $A$, whereas associations
$A\to B$ and $A\to C$, even together, provide
less information, as $B$ and $C$ could appear often
with $A$ but not so much together (we offer an example below). 
Thus, we do not force $|Y| = 1$.

In many cases we assume that the context provides for a
threshold on the confidence, imposing a constraint 
$c(X\to Y)\geq\gamma$ on rules, and likewise a support
threshold constraint $s(X\to Y)>\tau$. It is formally
convenient to use strict inequality in the latter case,
to easily cater for the case where no support bound is 
imposed, by simply taking $\tau=0$;
whereas, for confidence, we prefer to be able to
select full-confidence implications via the nonstrict
inequality with $\gamma = 1$. 


\begin{remark}
\label{rm:antcons}
As we consider mainly confidence and support, 
rules $X\to Y$ and $X\to XY$ are equivalent
in almost all our statements,
as are all rules where some part of the left-hand 
side $X$ is repeated in the right-hand side. 
Our novelty notions will respect as well this 
equivalence. The only exceptions will be in our
brief considerations of lift. Two natural canonical choices to
simplify the discussion are to restrict the discussion
either to
the rules of the
form $X\to Y$ or to those of the form $X\to XY$, where, 
in both cases,
$X\cap Y = \emptyset$.
We will see in Subsection~\ref{ss:mmr} that
failing to clarify this option risks overlooking
subtle differences among sets of rules enjoying,
however, quite different properties.
Based on the similar developments in implications
and functional dependencies, we choose the latter: 
we will make explicit always what part 
of the consequent is already in the antecedent and write all 
our association rules as $X\to XY$ where $X\cap Y = \emptyset$.
However, this choice is somewhat arbitrary, and whomever
prefers association rules with disjoint sides only needs
to remove the copy of the antecedent from the consequent.
In fact, in our implementations, at the time of showing a rule  
to the user, of course only the $Y$ part of the consequent is shown.
\end{remark}

Given a dataset $\D$, an itemset $X\subseteq\U$ is \emph{closed} if the support
of any strictly larger 
itemset
is strictly smaller; and is 
\emph{free}, or a
\emph{minimal generator}, if the support of any strictly smaller
itemset 
is strictly larger.
We denote as $\cl{X}$ the closure of itemset $X$ with respect to a given
dataset: $\cl{X}$ is the smallest closed itemset that includes $X$ or,
equivalently, the largest itemset that includes $X$ and has the same
support as $X$ in the dataset. It is easy to check that it is unique.
The intersection of closed itemsets is closed and, ordered by inclusion, 
the closed itemsets form a lattice which we call ``closure space''.
We will make liberal use of the three characteristic properties
of closure operators, namely, extensivity: $X\subseteq\cl{X}$; 
monotonicity: $X\subseteq Y$ implies $\cl{X}\subseteq\cl{Y}$; and
idempotency: $\cl{\cl{X}} = \cl{X}$.
We will mention below further details about the connections of
closure operators and free sets 
with association mining; see~e.~g.~\cite{BBR,Zaki} for
further information.

\begin{example}
\label{ex:running}
We will employ as running example through most of this paper
the closure space obtained from a specific dataset. For this
example, the universe $\U$ includes the five items $A$, $B$, $C$,
$D$, and $E$. The dataset consists of 12 transactions, six of which
include all of $\U$; two more consist of $ABC$, again two transactions
consist of $AB$, and then one transaction consists of $CDE$ and
another one consists of $BC$. It is easy to see that the 
associated closure-space lattice is as depicted in Figure~\ref{fg:clspace},
where transitive arcs have been omitted and, besides the closed sets,
three minimal generators (connected to their closures) have been indicated
in broken lines. The supports of all closed sets are reported
in the figure for convenience. The support of each minimal
generator coincides with that of its closure. Note that sometimes
the minimal generator coincides with its closure, as in set $BC$, 
for one.
This example illustrates that, at confidence 9/11, both the
association rules $B\to A$ and $B\to C$ hold, whereas the stronger
rule $B\to AC$ does not, as its confidence is only 8/11. That is,
if and when $B\to AC$ holds, it would give more information than
$B\to A$ and $B\to C$ holding jointly.
\end{example}


\begin{narrowfig}{0.5\textwidth}
\includegraphics[width=0.5\textwidth]{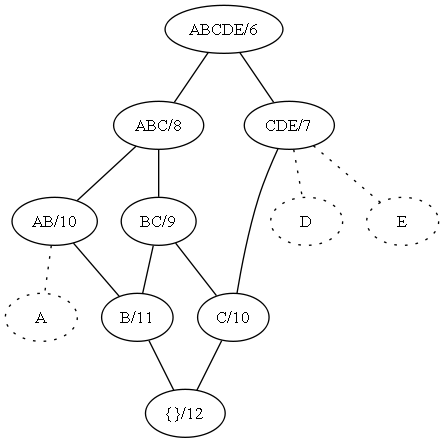}
\caption{An example closure space, with three minimal generators; the dataset 
contains the following transactions:
$ABCDE\,(\times6)$, 
$ABC\,(\times2)$, 
$AB\,(\times2)$, 
$CDE\,(\times1)$, 
$BC\,(\times1)$}
\label{fg:clspace}
\end{narrowfig}

We will propose to measure the novelty of each rule with respect 
to the rest of the outcome of the same data mining process, through
a variant of the intuitive idea of redundancy. Several notions
of redundancy for association rules exist. 
In the early proposal \cite{Lux}, a rule is redundant if
its confidence can be computed from that of other rules.
Later, this idea has been refined, making precise what 
information is maintained and which operations are allowed
to infer confidence or support of redundant rules:
see the survey of several concise
representations and redundancy notions in \cite{Krys02}.
In~\cite{PasBas} (and in earlier conference versions of
their work) the following set of rules is shown to be
sufficient to compute the confidence and support of any 
given partial rule:

\begin{definition}
\label{d:minmax}
Given a dataset and a support threshold $\tau$ acting
on all sets and rules:
\begin{enumerate}
\item
The \emph{min-max rules} are those of the form
$X\to XY$ where $XY$ is a closed set and $X$ is a minimal
generator; they are split into the following two cases.
\item
The \emph{min-max approximate rules} are those of the form
$X\to XY$ where $XY$ is a closed set, $X$ is a minimal
generator, and $\cl{X}\subset XY$.
They have confidence less than~1.
\item
The \emph{min-max exact rules} are those of the form
$X\to XY$ where $XY$ is a closed set, $X$ is a minimal
generator, and $\cl{X} = XY$.
They have confidence~1.
\end{enumerate}
\end{definition}


Similar notions of redundancy are studied in \cite{Zaki}, 
where, however, the approximate bases are constructed
as rules having minimal generators both at the left- and
at the right-hand sides. These bases are quite more
succinct than the sets of all association rules that
hold in a specific dataset, yet they still conform 
far too large quantities in many cases of interest.
Therefore, less demanding notions of redundancy 
for association rules have been studied. If we assume
that the set of frequent closures is kept, so that
confidences are easily computed from them, and focus
on the ``user-centric'' view, there is a very 
precise and natural notion that allows us to identify
irredundant bases of absolutely minimum size.
For the whole paper,
we concentrate basically on this redundancy notion,
and on a somewhat more sophisticate variant that we will
describe in Section~\ref{s:cbcb}.

\begin{lemma}
\label{l:redchar}
Consider two association rules, $X_0\to X_0Y_0$ and $X_1\to X_1Y_1$.
The following are equivalent:
\begin{enumerate}
\item The confidence and support of $X_0\to X_0Y_0$ 
are always larger than or equal to those of $X_1\to X_1Y_1$, 
in {\em all} datasets; that is, for every dataset $\D$, 
we will have $c_{\D}(X_0\to X_0Y_0)\geq c_{\D}(X_1\to X_1Y_1)$ and 
$s_{\D}(X_0Y_0)\geq s_{\D}(X_1Y_1)$ in it.
\item $X_1\subseteq X_0\subseteq X_0Y_0\subseteq X_1Y_1$.
\end{enumerate}
\end{lemma}

When these cases hold, we say that $X_1\to X_1Y_1$ makes 
$X_0\to X_0Y_0$ {\em redundant}, or also that $X_1\to X_1Y_1$ is 
{\em logically stronger} than $X_0\to X_0Y_0$. 
The notions come, essentially, 
from \cite{AgYu,KryszPAKDD}.
For a fixed confidence threshold, those rules that reach it, 
and are not made redundant by other rules also above the 
threshold, form the {\em representative} (or {\em essential}) 
{\em rule basis} for that confidence threshold 
\cite{AgYu,KryszPAKDD,PhanLuongICDM};
that is, every rule that reaches the confidence threshold
is either in the corresponding representative basis, 
or made redundant by a rule in the basis. 
Hence, a redundant rule is so because we can know 
beforehand, from the information in a basis, that its 
confidence will be above the threshold. 
These references also explain how to compute 
the representative basis out of the closed 
itemsets for the dataset.

The fact that statement 
\emph{(2)} implies statement \emph{(1)}
in Lemma~\ref{l:redchar} is easy to see 
and was already pointed out in \cite{AgYu,KryszPAKDD,PhanLuongICDM} 
(in somewhat different terms).
The 
converse implication
is nontrivial and much more recently 
shown~\cite{Bal10b}; see this reference 
as well for the proof 
that the representative basis has the 
minimum possible size among all bases for this notion
of redundancy, and for discussions of
other related redundancy notions.
In particular, several other natural proposals
are shown there to be equivalent to this redundancy.
Also, from this same source,
we will consider later on a variant 
which makes a deeper use 
of the closure operator., 

A 
known
property 
that relates
representative rules 
to
closure-based miners is:

\begin{proposition}
\label{p:rrclos}
On a given dataset and in the presence of a fixed support threshold~$\tau$,
consider the association rule $X\to XY$, and set $\gamma = c(X\to XY)$.
The following are equivalent:
\begin{enumerate}
\item
$X\to XY$ is a representative rule for some confidence threshold.
\item
$X\to XY$ is a min-max rule:
$XY$ is a closed set and $X$ is a minimal
generator.
\item
$X\to XY$ is a representative rule for confidence threshold $\gamma$.
\end{enumerate}
\end{proposition}

Hence, whenever we refer to $X\to XY$ as a
representative rule, without mention of the specific
confidence threshold $\gamma$ for which it is so, 
we implicitly understand that we mean $\gamma = c(X\to XY)$.
The implication from \emph{(1)}~to~\emph{(2)}~is from~\cite{KryszRSCTC}
(see also \cite{KryszIDA} for a clearer notation):
if $X\to XY$ is a representative rule then $s(X)<s(X')$
for all $X'\subset X$, and $s(Z)<s(XY)$ for all $Z$ with
$XY\subset Z$; that is, $X$ is a minimal generator and
$XY$ is closed. 

We have not found the other implications explicitly stated, 
but they appear implicitly, in a sense, in the references
that discuss these notions. We sketch here the rather simple 
proofs for completeness. 
Set $\gamma = c(X\to XY)$.
We assume that $X\to XY$ is a min-max rule, and
consider a different rule, $X'\to X'Y'$, logically 
stronger than $X\to XY$; we must argue that it fails
the confidence threshold. 
By Lemma~\ref{l:redchar}, we have
$X'\subseteq X$ and $XY\subseteq X'Y'$.
If the left-hand sides differ, $X'\subset X$ and,
$X$ being a minimal generator, $s(X') > s(X)$; then
$c(X'\to X'Y') \leq c(X'\to XY) < c(X\to XY) = \gamma$.
If, instead, $X' = X$, then $XY\subset X'Y'$ and,
$XY$ being closed, $s(X'Y') < s(XY)$; we obtain that
$c(X'\to X'Y') = c(X\to X'Y') < c(X\to XY) = \gamma$ again.
The remaining implication, \emph{(3)}~to~\emph{(1)}, is obvious.

\begin{example}
\label{ex:rrules}
One can check that the dataset and the closure space 
of Example~\ref{ex:running} lead to seven
representative rules at confidence threshold 0.8, namely,
$A\to BC$, $C\to AB$, $B\to C$, $\emptyset\to C$, $\emptyset\to AB$, and
$D\to ABCE$, and $E\to ABCD$.
The first two have confidence exactly 0.8, the others have confidences
slightly higher.
\end{example}

For fixed confidence thresholds, the representative rules
at that confidence 
form often a properly smaller basis
than the min-max rules; this
can be achieved because of two reasons.
One is that, obviously, min-max rules of
confidence below the threshold are omitted.
But a more sophisticate reason is that a
representative rule at a given confidence $\gamma$
may cease to be so at lower confidences:
at a lower threshold $\gamma'$ it is possible
that a stronger rule appears that makes it redundant.
This observation is key in the notion of confidence width
that we review next.

\subsection{Confidence Width}

Along most of our discussions in this paper, 
we assume that a dataset $\D$ and 
a support threshold $\tau$ have been fixed: all our rules
are assumed to reach strictly above that support threshold on $\D$. 

According to the definition of redundancy in Lemma~\ref{l:redchar},
all rules in the representative basis provide some irredundant
information. However, it is often the case that still the 
representative basis contains more rules than reasonable
for human inspection. 
In \cite{Bal09},
the intuition of redundancy is pushed further in order
to gain a perspective of novelty of association rules. 
An irredundant rule of a given confidence $c$ belongs to the basis 
for that confidence threshold $\gamma = c$: no rule of that confidence
or higher makes it redundant; equivalently, all rules that
make it redundant have lower confidence.
Then, one can ask: 
``how much lower?''. 
This can be evaluated by means of the following definition
from the same reference:


\begin{definition}
\label{d:width}
For an association rule $X\to XY$, 
consider all rules that are 
not equivalent to $X\to XY$ (as per Remark~\ref{rm:antcons}), 
but such that $X\to XY$ is 
redundant with respect to them, and pick one with maximum 
confidence in $\D$ among them, say $X'\to X'Y'$.
The {\em confidence width} of $X\to XY$ in $\D$
is: 
$$
w(X\to XY) = \frac{c(X\to XY)}{c(X'\to X'Y')}
$$
\end{definition}

The condition that $X\to XY$ is 
redundant with respect to $X'\to X'Y'$
implies that $c(X'\to X'Y')\leq c(X\to XY)$,
hence the confidence width is always~1 or larger.
In fact, $w(X\to XY)$ is strictly higher
than 1 if and only if $X\to XY$ 
is a representative rule.

To explain better the intuition behind the notion of 
confidence width, consider a rule $X\to XY$ of a given 
confidence, say $c(X\to XY) = c_0\in[0,1]$, and let us see 
what happens as we mine the representative basis at a 
varying confidence threshold~$\gamma$. 
If $c_0<\gamma$, the rule at hand will not play any role at all, being of 
confidence too low for the threshold. At $\gamma = c_0$, the rule becomes 
part of the output of any standard association mining process, but it could be
that some other ``logically stronger'' rule appears at the same 
confidence~$c_0$. 
For instance, it could be that both rules $A\to AB$ and $A\to ABC$ have 
confidence~$c_0$: then $A\to AB$ is redundant and will not belong to
the basis for that confidence. In this case, the confidence width
is~1, its smallest possible value.

If no stronger rule appears at threshold $\gamma = c_0$,
then $X\to XY$ will belong to the representative 
basis for that threshold. Let us keep decreasing the threshold. At some
lower confidence, a logically stronger rule may appear. If a logically 
stronger rule shows up early, at a confidence threshold $\gamma$ very close 
to $c_0$, then the rule $X\to XY$ is not very novel: it is too similar
to the logically stronger one, and this shows in the fact that the interval 
of confidence thresholds where it is a representative rule is narrow.
Its confidence width will be barely above~1.
To the contrary, a stronger rule may take long to appear: in that case, 
only rules of much lower confidence entail $X\to XY$, so that the fact that 
it does reach confidence~$c_0$ is novel in this sense. The interval of 
confidence thresholds where $X\to XY$ is a representative rule is wide,
as will be the value of the confidence width.
For instance, if the confidence of $A\to AB$ is 0.9, and all rules
that make it redundant have confidences below 0.75, the rule is a 
much better candidate to novelty than it would be if some rule like 
$A\to ABC$ would have a confidence of 0.88: in this last case,
$A\to AB$ 
indeed
brings in additional 
information, but its novelty, with respect to the 
other
rules,
is not high; it only belongs to the basis when
the confidence threshold is in the interval $(0.88,0.9]$. 
In the other case where all rules that could make it 
redundant have confidences, say, 0.75 or less,
then $A\to AB$ would belong to the basis for a 
considerably wider interval of confidences, $(0.75,0.9]$. 
It states something really different from the rest of the 
information mined. As an objective novelty measure, thus,
confidence width measures the width of the interval of
confidences in which the rule at hand belongs to the
representative basis.

It is proved in \cite{Bal09} that, in Definition~\ref{d:width},
it suffices to consider representative rules for the role of $X'\to X'Y'$.

\begin{example}
For association rule $A\to BC$, of confidence 0.8, 
in Example~\ref{ex:running}, 
the confidence width is 1.2. 
The confidence of that rule is 
at least 20\% higher than that of any rule that entails it.
Indeed, there are two representative rules
logically stronger, namely $A\to BCDE$ (of~confidence~0.6) 
and $\emptyset\to ABC$ (of higher confidence,~2/3);
hence, $w(A\to BC) = (8/10)/(2/3) = 1.2$.
\end{example}

Below we will need Definition~\ref{d:width} in a single formula; 
for this, we can replace the redundancy condition with 
its characterization according to Lemma~\ref{l:redchar}: 
$w(X\to XY) = {}$
$$
{} =
\frac{c(X\to XY)}{\max \{ c(X'\to X'Y') \st
(X\to XY) \neq (X'\to X'Y'), \, X'\subseteq X, \, XY\subseteq X'Y'\}}
$$
where again we are assuming that $X\cap Y = \emptyset$ and 
$X'\cap Y' = \emptyset$.

For each fixed support, there are rules that are not redundant 
with respect to any other, different rule; then, this quotient 
is undefined due to the emptiness of the set in the denominator,
for instance, if all candidate rules to it are of too low support.
By convention, we use $\infty$ as value of the confidence width 
in that case (equivalently, likening the max to a zero). We can 
identify easily which rules have infinite width (this proposition
is reported here for the first time):

\begin{proposition}
\label{p:infwidth}
The value of $w(X\to XY)$ is finite and well-defined if
and only if either $X\neq\emptyset$, or $Y$ has some 
proper superset $Z$ with $s(Z)>\tau$.
\end{proposition}

\noindent
{\sl Proof.}
Indeed, if $X=\emptyset$ and no proper superset of $Y$
reaches support above $\tau$ in the dataset, then 
no rule can make $\emptyset\to Y$ redundant; conversely, 
for $s(Z) > \tau$,
$\emptyset\to Z$ is different from $X\to XY$ and makes it
redundant if either $X\neq\emptyset$ and $Z = XY$, 
or $XY\subset Z$;
since this second case only needs to be applied
to rules with $X=\emptyset$, $Y\subset Z$ suffices.\qed

Thus, the only rules of infinite width are of the
form $\emptyset\to Z$ with $Z$ maximal under the
condition that $s(Z)>\tau$, and their confidence
would coincide with the normalized support of $Z$.
We observe in passing that, in practice, such 
maximal $Z$'s usually have a support barely above $\tau$,
because all supersets must have a support falling below~$\tau$;
whenever the confidence threshold is
substantially higher than 
the normalized support threshold
(which does not happen always but extremely often),
all rules of infinite width will be filtered out by
the confidence constraint.

It is easy to prove a simple observation,
that will be useful to compare below with confidence boost:
consider the condition $XY\subseteq X'Y'$ 
in the rules entering the maximization of the denominator; 
it can be written equivalently as follows, using the other 
condition that $X'\subseteq X$ and the empty-intersection 
assumptions: 

\begin{proposition}
\label{p:denom}
Assume $X'\subseteq X$, $X\cap Y = \emptyset$, and $X'\cap Y' = \emptyset$.
Then $XY\subseteq X'Y' \iff (X-X')\subseteq Y' \hbox{ and } Y\subseteq Y'$.
\end{proposition}

In \cite{Bal09}, some intuitions are described that suggest that, for
a confidence threshold $\gamma$, a natural choice could be to set the
confidence width threshold at $2-\gamma$; however, so far no formal
support for this proposal (or any other proposal, for that matter) is
known.

\subsection{Blocking Rules}
\label{ss:block}

On the basis of a clear, simple intuition described in many papers 
(e.g.~\cite{BAG,LiuHsuMa,PadTu2000,ShahLaksRS,ToKleRHM} 
just to name a few), \cite{Bal09} 
proposes also a notion of ``rule blocking'', 
whereby a subset of the antecedent may ``block'' an
association rule, that is, forbid its being provided
in the output, if the confidence of the rule with the smaller 
antecedent and the same consequent is higher enough.

The main question behind this option is the following.
Consider an association rule $X\to XY$, and 
reduce the antecedent to a smaller $Z\subset X$.
Whereas, intuitively, the rule with larger antecedent should
be subsumed by the other, this is due to the human intuitive habit
of working with full implications, where indeed this is the case.
But this is not so anymore with association rules.
For instance, at confidence 1, if $A\to C$ holds,
then $AB\to C$ also holds, and does not bring new information.
But 
association rules 
are not implications; 
instead, they relate relative frequencies:
compared to $X\to XY$, a smaller antecedent $Z\subset X$
does {\em not} lead to a new rule $Z\to ZY$ that entails it. 
Actually, for
$Z\subset X$, either of $X\to XY$ or $Z\to ZY$ may have
arbitrarily higher confidence than the other. 
Indeed: rule $X\to XY$ speaks about the abundancy of $Y$ among
the population of transactions that contain $X$;
reducing the antecedent into $Z$ changes the population
into, in principle, a larger one, and $Y$ can be
distributed at very different rates along each of these 
two sets of transactions. 
The distribution of $Y$ in the larger
population supporting $Z$ can be very imbalanced, 
so that $Y$ can appear more 
frequently in either.
%

\begin{example}
Consider two
association rules like $A\to C$ and $AB\to C$. 
It is easy to construct examples where 
almost all transactions with $A$ and $B$ have $C$, but 
they are a small fraction of those having $A$, and thus
the confidence of $A\to C$ is very small, whereas that of
$AB\to C$ is high, even~1; conversely, $C$ might hold for
nearly all of the transactions having $A$, 
but it could be that the only
transactions having both $A$ and $B$ are those without $C$ 
and, then, the confidence of $AB\to C$ can be zero yet
the confidence of $A\to C$ can be very high. 
\end{example}

\begin{example}
\label{ex:toblock}
Returning briefly to the dataset of Example~\ref{ex:running},
it is easy to check that 
$c(\emptyset\to BC) < c(A\to BC)$
whereas
$c(\emptyset\to C) > c(B\to C)$.
\end{example}

As a consequence, we also find the fundaments of the criticism
that confidence does not detect negative correlations.


\begin{example}
Fix a confidence threshold at 0.75, and consider a simple dataset 
with 10 transactions: 3 transactions $BC$, 6 transactions just $C$, 
and 1 transaction $B$. Then $c(B\to BC) = 0.75$, reaching the confidence 
threshold. Most association miners would report $B\to C$ as 
interesting at that threshold. However, the correlation 
between $B$ and $C$ is actually negative. Indeed, 
$C$ is \emph{less} frequent among the transactions having $B$ than 
in the total population, as $c(\emptyset\to C) = s(C)/n = 0.9$.
\end{example}

The natural reaction, consisting of a normalization by dividing 
the confidence by the (normalized) support of the consequent 
of the rule, gives a parameter that we find in the references
going by several different names: it has been 
called {\em interest} \cite{BrinMS} or, 
in a slightly different but fully equivalent form, 
{\em strength} \cite{ShahLaksRS}; 
``lift'' seems to be catching up as a short name, 
possibly aided by the fact that the Intelligent Miner 
system from IBM employed that name. 
The quantity is well-known in basic probability, 
as it measures the deviation from independence,
as a multiplicative distance from the
case of fully independent $X$ and $Y$, which would 
give value 1 for it: 

\begin{definition}
\label{def:lift}
The {\em lift} of rule $X\to Y$ is $\frac{c(X\to Y)}{s(Y)/n} =
\frac{s(XY)\times n}{s(X)\times s(Y)}$.
\end{definition}

(If supports are already normalized, then the factor $n$ for
the dataset size in the numerator has to be omitted.) 
The related parameter \emph{leverage}
\cite{PSG}
measures essentially the same thing, just that it does
so as an additive distance. 
It must be noted that, contrary to confidence, 
the lift of $X\to Y$ does not coincide with 
the lift of $X\to XY$: if we are to use lift, 
then we must be careful to keep the right-hand side
$Y$ disjoint from the left-hand side: $X\cap Y = \emptyset$.
Otherwise, misleadingly higher lift values are obtained.
Note also that, in case $X=\emptyset$, the lift
trivializes to~1.

However, this natural 
measure lacks the ability to orient the rules, because, 
in it, the roles of $X$~and~$Y$ are symmetric. 
Additionally, lift is limited in its ability to
control cases where $c(Z\to Y) > c(X\to Y)$
for $\emptyset\neq Z\subset X$. We describe a case
found in data from real census information,
pointed out also in \cite{Bal09}.
Mining for association rules at 5\% 
support~and~100\% confidence the \ds{Adult} dataset 
from Irvine \cite{UCI}, 67 (out of 71) rules in the 
basis are of the form 
``Husband'' + something else $\implies$ ``Male'', 
and the other four rules are also of this form except
for the addition of one more item in the consequent. 
The reason is that the rule 
``Husband'' $\to$ ``Male'',
that we would expect to hold, does not reach 100\% confidence: 
indeed, tuple 7110 includes the items ``Husband'' and ``Female'' 
(instead of ``Male''). This opens the door to many rules,
intuitively uninformative, that enlarge a bit the left-hand side, 
just enough to avoid tuple 7110 so as to reach confidence 100\%. 
The whole issue would not be solved by dividing all confidences
by the support of ``Male''. 
Further examples are given in the same paper, and in many 
others such as those cited above.



It is desirable to react to the negative correlation problem
for confidence and still maintain orientability.
As an alternative approach to this problem, in \cite{Bal09} 
the confidence parameter is used in an intuitive way to find a
threshold at which a 
smaller antecedent would suggest to omit
a given rule. The proposal there is 
fully equivalent to the following one: 

\begin{definition}
Given rule $X\to XY$, with $X\cap Y = \emptyset$, 
a proper subset $Z\subset X$ {\em blocks} $X\to XY$ 
at blocking threshold $b$
if 
$$
\frac{s(XY) - c(Z\to ZY)s(X)}{c(Z\to ZY)s(X)} \leq b.
$$
\end{definition}

The threshold $b$ is intended to take positive but
small values, say around 0.2 or lower.
The intuition behind this definition is as follows:
we will want to discard rule $X\implies XY$ in case we find 
a rule $Z\implies ZY$, with $Z\subset X$ (and
therefore $ZY\subset XY$, also properly), having 
``almost'' the same confidence, or larger.
(In the presence of a support threshold $\tau$,
we would be requiring as well, naturally, that
$s(Z\implies ZY) > \tau$.)
To do this, we compare the number of tuples having $XY$ 
with the quantity that would be predicted from the confidence 
of the rule $Z\implies ZY$. 

More precisely, let $c(Z\implies ZY) = c$. If $Y$ is
distributed along the support of $X$ at the same ratio as along 
the larger support of $Z$, we would expect $s(XY)\approx c \times s(X)$:
we are, thus, considering the relative error committed by
$c\times s(X)$ used as an approximation to $s(XY)$.
In case the difference in the numerator is negative, it would mean 
that $s(XY)$ is even lower than what $Z\implies ZY$ would suggest. 
If it is positive but the quotient is low, $c(Z\implies ZY)\times s(X)$
still suggests a good approximation to $c(X\implies XY)$, and
the larger rule
does not bring high enough confidence with respect to 
the simpler one
to be considered: it remains blocked. But, if the quotient is larger, 
and this happens for all $Z$, then $X\implies XY$ becomes interesting 
since its confidence is higher enough than suggested by other rules of 
the form $Z\implies ZY$. 

The higher the block threshold, the more demanding 
the constraint is. It can be checked that the particular problems 
of the \ds{Adult} dataset indicated above are actually solved 
already by imposing just a generously tiny blocking threshold
(around 0.000075).
Again the specific choice of value for the blocking
threshold is justified in~\cite{Bal09}
just in merely intuitive terms; however, note for later use that the
confidence width bound and the blocking threshold are related in that
paper as follows: if the confidence width bound is $b$, then the 
blocking threshold proposed is $b-1$.

\begin{example}
\label{ex:block}
Due to the inequalities in Example~\ref{ex:toblock}, we can see that,
at any nonnegative blocking threshold, $\emptyset$ blocks $B\to C$:
$$
\frac{s(XY) - c(Z\to ZY)s(X)}{c(Z\to ZY)s(X)} 
=
\frac{s(BC) - c(\emptyset\to C)s(B)}{c(\emptyset\to C)s(B)} 
\approx
\frac{9 - 9.16}{9.16} < 0.
$$
Likewise, considering $A\to BC$, we have
$$
\frac{s(ABC) - c(\emptyset\to BC)s(A)}{c(\emptyset\to BC)s(A)} 
=
\frac{8 - (9/12)*10}{(9/12)*10} 
\approx
0.066
$$
so that this rule would be blocked by $\emptyset$
as soon as a blocking threshold 
higher than this quantity is imposed.
\end{example}

\subsection{Support Ratio}
\label{ss:suppratio}

We will relate our values of confidence width and of confidence
boost to an expression essentially employed first, to our knowledge, in
\cite{KryszIDA}, where no particular name was assigned to it.
Together with other similar quotients, it was introduced with
the aim of providing a faster algorithm for computing 
representative rules; it turns out that, as demonstrated
in \cite{BalTir11a}, this approach is efficient and useful
in practice but runs into the risk of providing incomplete
output, as actual representative rules may be missed.
The same reference 
provides further analysis,
including almost equally efficient alternatives whose output
is complete.

Here we introduce this notion because 
it is related to all of our three parameters
of confidence width, blocking, and confidence boost;
it will allow us to explain more carefully their 
mutual relationships, and it allows for confidence
boost constraints to be ``pushed'' into a closure
mining process, as we will do in Section~\ref{s:yacaree}.

\begin{definition}
\label{d:suppratio}
In the presence of a support threshold $\tau$, 
the {\em support ratio} of an association rule $X\to XY$ 
is 
$$
\sigma(X\to XY)
 = 
\frac{s(XY)}{\max \{ s(Z) \st XY\subset Z, \, s(Z) > \tau \}}
$$
\end{definition}

We see that this measure 
does not depend on the antecedent~$X$ but just on $XY$. Again, 
we set its value to $\infty$ if no $Z$ exists as required for the 
maximization in the denominator. We have the following relationship:

\begin{proposition}
\label{p:suppratiowidth}
If the value of $\sigma(X\to XY)$ is finite and well-defined 
then the confidence width $w(X\to XY)$ is also finite, and
then 
$$
w(X\to XY) \leq \sigma(X\to XY).
$$
\end{proposition}

\noindent
{\sl Proof.}
This is easy to see from Proposition~\ref{p:infwidth},
and by observing that $X\to Z$, for the $Z\neq XY$ that 
maximizes the support in the denominator of support ratio, 
leads to 
$w(X\to XY) \leq c(X\to XY)/c(X\to Z) = s(XY)/s(Z) = \sigma(X\to XY)$
by simplifying the value of $s(X) \neq 0$.\qed

It is clear that $\sigma(X\to XY)\geq 1$ for all rules;
$\sigma(X\to XY) = 1$ exactly when $XY$ is not closed, 
since these sets are those that have some proper 
superset $Z$ with the same support.
The following easy consequence is worth mentioning: many of
the quantities we study for an association rule $X\to XY$ 
are bounded from above by the support ratio and, therefore,
will trivialize to values less than or equal to 1 unless 
we consider only closed sets $XY$ as right hand sides. 
Together with Proposition~\ref{p:rrclos}, this is the reason 
of the importance of the closure notion in our context, and 
of the introduction of a closure-aware version of confidence boost 
in Section~\ref{s:cbcb}.

\begin{example}
Looking again at association rule $A\to BC$
in Example~\ref{ex:running}, we see that
$\sigma(A\to BC) = s(ABC)/s(ABCDE) = 4/3$.
\end{example}

\section{Confidence Boost}
\label{s:cb}

This section introduces the first, simpler version of our main 
notion; it is very similar to the one given for confidence width, 
but with a twist that, even though formally tiny, semantically 
changes it far enough so as to encompass the notion of blocking.

\begin{definition}
\label{d:boost}
The {\em confidence boost} of an association rule $X\to XY$ 
(always with $X\cap Y = \emptyset$) is $\beta(X\to XY) = {}$
$$
{} = 
\frac{c(X\to XY)}{\max \{ c(X'\to X'Y') \st
(X\to XY) \neq (X'\to X'Y'), \, X'\subseteq X, \, Y\subseteq Y'\}}
$$
\end{definition}

As in previous cases, the rules in the denominator
are implicitly required to clear the support threshold:
$s(X'\to X'Y') > \tau$.
Again, in case the set in the denominator is empty,
the confidence boost is infinite by convention.
As in Proposition~\ref{p:infwidth}, we can point out
exactly which rules fall in that case: the same ones,
in fact.

\begin{proposition}
\label{p:infboost}
The value of $\beta(X\to XY)$ is finite and well-defined if
and only if either $X\neq\emptyset$, or $Y$~has some 
proper superset $Z$ with $s(Z)>\tau$. That is: the set of 
rules of infinite confidence boost coincides with the set
of rules of infinite width.
\end{proposition}

\noindent
{\sl Proof.}
Like in Proposition~\ref{p:infwidth}, 
if $X=\emptyset$ and no proper superset of $Y$
reaches support above $\tau$ in the dataset, then 
no different rule (of sufficient support) is
available for the set in the denominator.
Conversely, $\emptyset\to Z$ belongs to that set
if either $X\neq\emptyset$, or $Y\subset Z$.\qed

As indicated above, these cases of infinite
confidence boost hardly ever appear in
practice, due to their confidence being below
the threshold.

\begin{example}
Considering again association rule $A\to BC$
in Example~\ref{ex:running}, we find a value
of the confidence boost of 16/15 for this rule.
This is obtained as follows: we consider all rules
$X'\to X'Y'$ with $X'\subseteq A$ and $BC\subseteq Y'$
(and different from it); one can see that the maximum
confidence among them is 0.75, attained by $\emptyset\to BC$.
Then $\beta(A\to BC) = 0.8/0.75 = 16/15 \approx 1.066$.
\end{example}

The fact that a low confidence boost 
corresponds to a low novelty is similar
to the analogous explanation for width,
and can be argued intuitively as follows.
Suppose that $\beta(X\to XY)$ is low, say
$\beta(X\to XY)\leq b$, where $b$ is just slightly larger than~1. 
Then,
according to the definition, there must exist 
some {\em different} rule $X'\to X'Y'$, with $X'\subseteq X$ and 
$Y\subseteq X'Y'$, such that $\frac{c(X\to XY)}{c(X'\to X'Y')}\leq b$, 
or $c(X'\to X'Y')\geq c(X\to XY)/b$. 
This inequality says that the rule $X'\to X'Y'$, stating that transactions 
with $X'$ tend to have $X'Y'$, has a confidence relatively high, not much 
lower than that of $X\to XY$; equivalently, the confidence of $X\to XY$ 
is not much higher (it could be lower) than that of $X'\to X'Y'$. But 
all transactions having $X$ do have $X'$, and all transactions having $Y'$ 
have $Y$, so that the confidence found for $X\to XY$ is not really that 
novel, given that it does not give so much additional confidence over a
rule that states such a similarly confident, and intuitively stronger, 
fact, namely $X'\to X'Y'$.

At a bare minimum, we should not consider association rules with
confidence boost~1 or less. Notice that this solves the objection 
against confidence that negative correlations go undetected:
for instance, if the support of $B$ is, say, 80\%, a rule $A\to B$ of 
confidence less than that would yield a confidence boost below~1,
due to the rule $\emptyset\to B$.

\subsection{Boost, Lift, Support Ratio, Width, and Blocking}
\label{ss:bwb}

We present now some analyses clarifying the properties
of the confidence boost. First, we see that it allows one
to filter out rules that would be discarded on the basis 
of lift, since rules of low lift have low confidence boost.

\begin{proposition}
\label{p:blift}
Let $X\neq\emptyset$; then, the confidence boost 
$\beta(X\to XY)$ is bounded from above by 
the lift of $X\to Y$.
\end{proposition}

\noindent
{\sl Proof.}
We simply consider the rule $\emptyset\to Y$, which
differs from $X\to Y$ since $X\neq\emptyset$. 
Its support is above that of $X\to Y$ and thus
above the support threshold. Clearly,
it appears among the rules considered to maximize
the confidence in the denominator of the definition
of $\beta(X\to XY)$, hence 
$\beta(X\to XY)\leq\frac{c(X\to XY)}{c(\emptyset\to Y)}$; 
but $c(\emptyset\to Y) = s(Y)/n$
and then $\frac{c(X\to XY)}{c(\emptyset\to Y)}$ is
exactly the lift of $X\to Y$.\qed

In the case where $X=\emptyset$, the lift is~1, 
as already indicated; this value turns out to be
uninformative in this case, since any right-hand side
is independent from~$\emptyset$. 
Confidence boost does apply to this case, being
able to detect low novelty through larger consequents.

The only formal difference between confidence boost
and confidence width of a rule $X\to XY$ is that, 
upon exploring alternative rules $X'\to X'Y'$, 
in the confidence boost the antecedent $X$ is {\em not} 
required anymore to be a subset of the consequent $X'Y'$, 
whereas it must be for $X'\to Y'$ to qualify in the
computation of the width.
More precisely, given that $X\cap Y = \emptyset$ and 
$X'\subseteq X$, it follows $X'\cap Y = \emptyset$,
so that the condition $Y\subseteq Y'$ is equivalent to 
the condition $Y\subseteq X'Y'$. 
Proposition~\ref{p:denom} tells us that
$XY\subseteq X'Y' \iff (X-X')\subseteq Y' \hbox{ and } Y\subseteq Y'$,
and we see that 
confidence boost simply keeps the inclusion among the right-hand
sides $Y\subseteq Y'$ 
and does not require additionally that $(X-X')\subseteq Y'$
anymore. 
This also tells us that all rules $X'\to X'Y'$ that are 
considered for the maximization in the denominator in
the definition of confidence width are also 
considered for the corresponding maximization 
in confidence boost. Thus, the value of the maximum itself is 
at least the same, or possibly larger, and the difference
is that the boost case may consider further candidates to $X'\to X'Y'$.
That is:

\begin{proposition}
\label{p:widthboost}
The confidence boost of a 
rule 
is bounded above by its confidence width:
$\beta(X\to XY)\leq w(X\to XY)$. Hence,
$\beta(X\to XY)\leq \sigma(X\to XY)$.
\end{proposition}

The last sentence comes from Proposition~\ref{p:suppratiowidth},
and was proved directly first in \cite{BalTirZor10a}.
For the next theorem, we state separately
a simple technical equivalence.

\begin{lemma}
\label{l:block}
$Z\subset X$ blocks $X\to XY$ at block 
threshold $b-1$
if and only if $\frac{c(X\to XY)}{c(Z\implies ZY)} \leq b$.
\end{lemma}

\noindent
{\sl Proof.}
By definition, $Z\subset X$ blocks $X\to XY$ at blocking threshold $b-1$
if and only if
$$
\frac{s(XY) - c(Z\implies ZY)s(X)}{c(Z\implies ZY)s(X)} \leq b-1.
$$
Multiplying both sides of the inequality
by $c(Z\implies ZY)$, separating the two
terms of the left-hand side, and replacing $s(XY)/s(X)$ by its
meaning, $c(X\to XY)$, we find the equivalent expression
$$
c(X\to XY) - c(Z\implies ZY) \leq (b-1)(c(Z\implies ZY)
$$
where solving for $b$ leads to
$$
\frac{c(X\to XY)}{c(Z\implies ZY)} \leq b.
$$
All the algebraic manipulations are reversible
(in particular, confidences and supports appearing all along
are never zero so we can multiply or divide by them without trouble.)\qed

We show next that confidence boost embodies exactly
both blocking and confidence width, precisely with the
same relation between the thresholds as used in \cite{Bal09},
under the already stated proviso that all the association rules
involved must clear the support threshold.

\begin{theorem}
\label{th:char}
For an association rule $X\to XY$,
$\beta(X\to XY) \leq b$ if and only if
either
$w(X\to XY)\leq b$ or 
$X\to XY$ is blocked at a blocking threshold $b-1$.
\end{theorem}

\noindent
{\sl Proof.}
First we prove that either of low width or blocking
imply low boost. 
We have already argued in Proposition~\ref{p:widthboost} that
$
\beta(X\to XY) \leq w(X\to XY)
$.
Likewise,
assume that $Z\subset X$ (proper subset)
blocks 
$X\to XY$ at a blocking threshold $b-1$. Clearly the rule
$Z\to ZY$ differs from $X\to XY$ since $Z$ is a proper subset of $X$
and fulfills the conditions to enter the maximum confidence
denominator in the definition of confidence boost. This means
that this maximum is at least as large as $c(Z\to ZY)$, and
therefore, by Lemma~\ref{l:block}, 
$$
\beta(X\to XY) \leq \frac{c(X\to XY)}{c(Z\implies ZY)} \leq b.
$$

Conversely, we assume now that 
$\beta(X\to XY) \leq b$ and prove that either 
$w(X\to XY)\leq b$ or 
$X\to XY$ is blocked at a blocking threshold $b-1$.
The definition of confidence boost tells us that
there is a different rule $X'\to X'Y'$ ($X'\cap Y' = \emptyset$)
for which $s(X'Y') > \tau$,
$X'\subseteq X$, $Y\subseteq Y'$, and 
$\frac{c(X\to XY)}{c(X'\implies X'Y')} \leq b$.
We consider two cases, according to whether
$X=X'$. If $X=X'$, necessarily $Y\subset Y'$
properly, thus $XY\subset X'Y'$ properly, and
$s(X) = s(X')$ plus $s(X'Y') > \tau$ tells us
that 
$$
w(X\to XY) \leq \sigma(X\to XY) \leq \frac{s(XY)}{s(X'Y')}
= \frac{c(X\to XY)}{c(X'\to X'Y')} \leq b.
$$
Otherwise, $X'\subset X$ properly, and $Y\subseteq Y'$
(and {\em a fortiori} $X'\cap Y = \emptyset$)
gives us $c(X'\to X'Y) \geq c(X'\to X'Y')$ whence
$\frac{c(X\to XY)}{c(X'\implies X'Y)} \leq 
\frac{c(X\to XY)}{c(X'\implies X'Y')} \leq b$.
Applying again Lemma~\ref{l:block}, we obtain
that $X'$ blocks $X\to Y$ at blocking threshold $b-1$.\qed

Hence, bounding the confidence boost at $b$ ensures us 
that the rules that would be filtered by 
that confidence boost bound
are exactly the same as those that would be filtered 
by either (or both) 
of the checks $w(X\to XY)\leq b$ or blocking at threshold $b-1$. 
In this
sense, confidence boost embodies both low-novelty tests 
from~\cite{Bal09}, and with the same thresholds employed there.

We briefly consider the case of rules with a single item
in the antecedent.

\begin{proposition}
\label{p:liftboostsingle}
Assume that 
$|X| = 1$ in rule 
$X\to XY$, that is, the left hand side is a single item.
Then $\beta(X\to XY)$ coincides with the minimum among
the lift of $X\to Y$ and $\sigma(X\to XY)$.
\end{proposition}

\noindent
{\sl Proof.}
Let $X'\to X'Y'$ be the rule that 
leads to $\beta(X\to XY) = c(X\to XY)/c(X'\to X'Y')$.
It must be different from $X\to XY$, and must clear
the support threshold.

If $X'\subset X$, as $X$ is a singleton, we have 
$X'=\emptyset$,
$s(X') = n$ 
(the number of transactions in the dataset), 
$Y \subseteq Y'$, $s(Y')\leq s(Y)$, 
and
$$
\beta(X\to XY) = \frac{c(X\to XY)}{c(X'\to X'Y')} =
\frac{c(X\to Y)}{s(Y')/n} \geq
\frac{c(X\to Y)}{s(Y)/n} =
\frac{s(XY)\times n}{s(X)\times s(Y)}
$$
which is the value of the lift; but the boost is also less
than or equal to the lift by Proposition \ref{p:blift},
and they must coincide. The support ratio must
be higher by Proposition~\ref{p:widthboost}, so the confidence
boost equals the stated minimum.

The other case is where $X' = X$; then, as the two
association rules are different, necessarily $XY\neq X'Y' = XY'$,
so that $\sigma(X\to XY) \leq s(XY)/s(XY') = c(X\to XY)/c(X\to XY')$
because we can divide by $s(X)\neq 0$; that is, 
$\sigma(X\to XY)\leq\beta(X\to XY)$. The converse inequality
is furnished by Proposition~\ref{p:widthboost} and, once we have
the equality $\sigma(X\to XY) = \beta(X\to XY)$, the fact that
this value is the indicated minimum comes from 
Proposition~\ref{p:blift}.\qed

\begin{corollary}
\label{c:quasitrackboost}
Assume a threshold $b$ in place such that $\sigma(X\to XY)\geq b$ is
known, for $|X|=1$, that is, for a rule with a single antecedent item.
If the lift of $X\to Y$ is less than $b$, then it equals
$\beta(X\to XY)$.
\end{corollary}

\begin{example}
\label{ex:properineqs}
We revisit again association rule $A\to BC$
in Example~\ref{ex:running}. For this rule,
the lift is 16/15, less than the support ratio 4/3,
so that the former coincides with the confidence boost
as per Proposition~\ref{p:liftboostsingle}. The quantities
evaluated in previous examples lead now to the inequalities
$$
\beta(A\to BC) = 16/15 < w(A\to BC) = 6/5 < \sigma(A\to BC) = 4/3
$$
which obey, of course, all inequalities we have
proved so far and, at the same time, witness that
each inequality may well be proper.
\end{example}

\subsection{Double-Threshold Confidence}
\label{ss:algo}

In order to be of practical use, we need a deeper study of the 
confidence boost. As it currently stands, it makes no sense to 
traverse all the alternative rules to be taken into account for
computing the maximum confidence in the denominator. 
The same 
sort of difficulty appears for confidence width and for blocking.
A mild precomputation allows one to compute quite efficiently the 
width \cite{Bal09},
but the same method does not seem to work for blocking or boost.
In fact, the experiments reported in that reference 
resort, as indicated there, to an approximation to blocking.

By the reasons already discussed, we will not be interested
in confidence boost bounds of 1 or less; above 1, by 
Proposition~\ref{p:widthboost}, 
we only find representative
rules. Given confidence threshold $\gamma$,
we will show that, in order to test the confidence boost
threshold, it suffices to do so against the set of representative
rules computed at a lower confidence threshold, namely $\gamma/b$.
Indeed, consider Algorithm~\ref{a:cb}.
The comparisons are written there in such a way so as to avoid division 
by zero in the cases of infinite boost, such as $s(XAY)=0$,
which may potentially be the case.

\begin{algorithm}
\label{a:cb}
\DontPrintSemicolon
\KwData{dataset $\D$;
thresholds for support $\tau$, 
for confidence $\gamma$, 
and for confidence boost $b>1$;
rule $X\to XY$ with $X\cap Y=\emptyset$, 
$c(X\to XY)\geq \gamma$, 
and $s(XY)\geq \tau$}
\KwResult{boolean value indicating whether $\beta(X\to XY) > b$}
mine $\D$ for the representative rules $\R$ at threshold $\gamma/b$\;
\For{each rule $X'\to X'Y'\in\R$ such that $X'\cap Y'=\emptyset$,
$X'\subseteq X$ and $Y\subseteq Y'$}{
\If{$\exists Z\subset X-X'$ such that $c(X\to XY)\leq b\times c(X'Z\to X'ZY)$}{
\Return{{\tt False}}}
\If{$\exists A \in Y'-XY$ such that $c(X\to XY)\leq b\times c(X\to XAY)$}{
\Return{{\tt False}}}}
\Return{{\tt True}}
\caption{A double confidence threshold algorithm}\label{a:dconf}
\end{algorithm}

%
%
%
%
%
%
%
%
%
%
%
%
%
%
%

\begin{theorem}
\label{th:algo}
Let $X\to XY$ be a rule of confidence at least $\gamma$. 
Then, Algorithm~\ref{a:dconf} accepts it if and only if $\beta(X\to XY) > b$.
\end{theorem}

\noindent
{\sl Proof.}
First we see that the rejections are correct. In each case, we just found
a rule $X''\to X''Y''$ with $X''\subseteq X$ and $Y\subseteq Y''$,
be it $X'Z\to X'ZY$ or $X\to XAY$; also $X''\to X''Y''\neq X\to XY$: 
in the first case, $Z$ is a proper subset
of $X-X'$, so $X'Z\neq X$, and in the second case the item $A$ did not
appear in $X\to XY$. In each case, the rule $X''\to X''Y''$ enters the
maximization in the denominator of the confidence boost and shows that
its value is less than or equal to~$b$. 

To see that acceptance is correct, assume $\beta(X\to XY)\leq b$: we
prove that, at some point, rule $X\to XY$ must fail one of the two tests
in the algorithm. By the definition of confidence boost, there must exist 
some rule $X''\to X''Y''$, different from $X\to XY$, 
with $X''\subseteq X$ and $Y\subseteq Y''$,
such that $c(X\to XY)\leq b\times c(X''\to X''Y'')$.

Then,
from $c(X\to XY)\geq \gamma$ we infer 
$c(X''\to X''Y'')\geq \gamma/b$, so that
there must exist a representative rule at confidence $\gamma/b$, 
let it be $X'\to X'Y'\in\R$, that makes $X''\to X''Y''$ redundant
(possibly itself): by Lemma~\ref{l:redchar}, $X'\subseteq X''$ and
$X''Y''\subseteq X'Y'$. At some point (unless a correct
negative answer is found earlier), the algorithm will consider 
this rule $X'\to X'Y'\in\R$. 
As in the proof of Theorem~\ref{th:char}, we distinguish two cases.

First assume that $X''$ is a proper subset of $X$, $X''\subset X$.
Since $X'\subseteq X''$, we can consider $Z=X''-X'\subset X-X'$:
at some point, the algorithm will compare $c(X\to XY)$ to 
$b\times c(X'Z\to X'ZY)$. But it holds that $X'Z = X''$ and that 
$Y\subseteq Y''$, resulting in 
$c(X\to XY)\leq b\times c(X''\to X''Y'') \leq b\times c(X'Z\to X'ZY)$
and failing the test.

Alternatively, assume $X''\subseteq X$ holds with equality: 
$X'' = X$. From $X''\to X''Y''\neq X\to XY$ (and using 
$X\cap Y=\emptyset$ and $X''\cap Y''=\emptyset$) we know that
$Y\subset Y''$ is a proper inclusion: there is some 
$A\in Y''\subseteq X'Y'$ that is not in~$Y$. Such $A$ is not
in $X$ either, because $X''\cap Y''=X\cap Y''=\emptyset$,
and then, in fact, $A\notin X'$, so that $A \in Y'-XY$.
In due time, the algorithm will compare $c(X\to XY)$ 
to $b\times c(X\to XAY)$. But $X=X''$, and $A\in Y''$ so that 
$AY\subseteq Y''$, hence   
$c(X\to XY)\leq b\times c(X''\to X''Y'') \leq b\times c(X\to XAY)$
and the test will fail as well. This completes the proof.\qed






\section{Closure-Based Confidence Boost}
\label{s:cbcb}

Representative rules are a minimum size basis for redundancy,
defined as per Lemma~\ref{l:redchar}; still, they constitute 
often a large set. Prior to accepting the option of losing 
information in a quantifiable manner, as we are doing via
confidence boost, one could consider the option of using stronger
notions of redundancy. 
Several earlier papers, e.~g.~\cite{Lux,PasBas,Zaki},
suggest to treat separately the implications, which allow
for more compact bases, from the partial rules.
In~\cite{Bal10b}, 
besides another more complicated alternative,
we follow up this suggestion as well, 
and employ a notion of closure-based
redundancy which also turns out to provide a complete basis 
of provably minimum size, denoted $\Bst{}$. This option has
definite advantages: 
whereas it provides bases comparable
in size with, and often clearly smaller than, 
the set of representative rules, it has the 
desirable property that the portion of it
that refers to partial associations (of confidence
below~1) can be computed faster. 
The best approaches to the representative rules
need to work on the basis of both the closures
lattice plus all the minimal generators of each
closure (\cite{KryszIDA}, but see the related
discussion in~\cite{BalTir11a});
instead, the $\Bst{}$ basis can be computed
just from the closures. 
In this section,
we port confidence boost into closure-based redundancy and
the corresponding minimum-size basis $\Bst{}$.

Closure-based redundancy corresponds to 
restricting consideration of datasets 
as a function of the closure operator they induce. It is
well-known that the closure operator is equivalently specified
by a set of implications, that is, association rules of 
confidence~1 (see e.~g.~\cite{Zaki}). Closure-based redundancy \cite{Bal10b}
takes into account the closure operator indirectly as follows:

\begin{definition}
Let $\B$ be a set of implications.
Partial rule $X_0\implies X_0Y_0$ has {\em closure-based
redundancy relative to $\B$} with respect to rule 
$X_1\implies X_1Y_1$ if the inequalities
$$
c(X_0\to X_0Y_0) \geq c(X_1\to X_1Y_1) \hbox{\quad and\quad} 
s(X_0\to X_0Y_0) \geq s(X_1\to X_1Y_1)
$$
hold in any dataset $\D$ in which all the rules in $\B$ 
hold with confidence~1.
\label{d:closredundancy}
\end{definition}

This redundancy has a characterization parallel to that of
Lemma~\ref{l:redchar}, proved in the same reference:

\begin{lemma}
\label{l:clredchar}
Let $\B$ be a set of implications.
Consider two association rules, $X_0\to X_0Y_0$ and $X_1\to X_1Y_1$.
The following are equivalent:
\begin{enumerate}
\item 
Rule $X_0\implies X_0Y_0$ has closure-based
redundancy relative to $\B$ with respect to rule 
$X_1\implies X_1Y_1$.
\item 
$X_1\subseteq\cl{X_0}$ and $X_0Y_0\subseteq\cl{X_1Y_1}$.
\end{enumerate}
The closure operator in the second statement is the one
corresponding to the set of implications $\B$.
\end{lemma}

In all applications, $\B$ is the set of full-confidence
implications holding in the dataset, so that the closure
operator is actually the one induced by the dataset.
For closure-based redundancy, a minimum-size basis can be constructed
as well. Essentially, this basis, denoted $\Bst{\gamma}$ for confidence
threshold $\gamma$, is defined in a manner analogous to that of the
representative rules, except that it is restricted to rules of the
form $X\to XY$ where {\em both} $X$ and $XY$ are closed sets,
instead of $X$ being a minimal generator as in representative rules. 
All these definitions are studied in depth in~\cite{Bal10b}. 

If we are to employ this notion of redundancy and the $\Bst{}$ basis,
then the definition of confidence boost requires some fine tuning. 
This basis is often smallish because many different representative rules
could correspond to many left-hand sides that are minimal
generators of the same closure. Such sets of rules become
a single rule in $\Bst{}$. But, if we use the given definition
of confidence boost, these rules are syntactically different from
the one in $\Bst{}$ and ``kill~it'' by forcing its boost down to~1.
Thus, to avoid trivializing $\Bst{}$, we need to take into account the 
closure operator in the definition of boost.
The main notion of this section is as follows:

\begin{definition}
\label{d:clboost}
The {\em closure-based confidence boost} of a rule $X\to XY$ 
is $\cl{\beta}(X\to XY) = {}$
$$
{} = 
\frac{c(X\to XY)}{\max \{ c(X'\to X'Y')^{\strut} \st
(\cl{X}\neq\cl{X'} \lor \cl{XY}\neq\cl{X'Y'}),
\, X'\subseteq \cl{X}, \, Y\subseteq \cl{X'Y'}\}}
$$
\end{definition}

This is the natural definition paralleling the confidence boost
when the notion of reduncancy is closure-based: on one hand, the
rules in the denominator may resort to the use of closures to make
the rule at hand redundant, widening the options of redundancy;
on the other hand, rules that are syntactically different from
the rule at hand, but equivalent to it in closure-based
redundancy, must be discarded, as they trivially entail 
the rule at hand.
Failing to discard them unduly trivializes the confidence boost
in many cases.
Observe that the notion of confidence boost in the previous
section 
corresponds to the particular case where the closure operator
is the identity function.

\begin{example}
\label{ex:cbcb}
Out of the seven representative rules at confidence threshold 0.8
that we enumerated in Example~\ref{ex:rrules}, some are unchanged
in $\Bst{0.8}$, such as 
$C\to AB$, $B\to C$, $\emptyset\to C$, and $\emptyset\to AB$.
Instead of $A\to BC$, we find $AB\to C$, which is equivalent to
it due to the implication $A\to B$; and, due to the implications
$D\to CE$ and $E\to CD$, it suffices to keep $CDE\to AB$ instead
of the other two.
If we were to employ plain confidence boost, 
$\beta(CDE\to AB) \leq 1$, due to rules $D\to ABCE$ and $E\to ABCD$.
Closure-based confidence boost is able to perform
a finer distinction. As these two rules have the same closure
of the antecedent as $\cl{D} = \cl{E} = CDE$, and the same
associated closed set $ABCDE$, they do not enter the computation
of closure-based confidence boost of $CDE\to AB$, which is
actually $\cl{\beta}(CDE\to AB) = c(CDE\to AB)/c(C\to ABDE) = 10/7 > 1$.
\end{example}

\subsection{Double-Threshold Confidence Revisited}
\label{ss:algocbcb}

We develop next an algorithm to compute
closure-based confidence boost. We just
need to make a number of adjustments to
the one given for plain confidence boost: first, one must
explore the rules of the $\Bst{}$ basis for confidence $\gamma/b$,
instead of the representative rules for it, since that is
the appropriate basis for closure-based redundancy; and,
second, one must take into account the closure 
operator at the time of checking whether a specific $\Bst{}$
rule may lead to guaranteeing low boost of the input rule.

\begin{algorithm}
\DontPrintSemicolon
\KwData{dataset $\D$;
thresholds for support $\tau$, 
for confidence $\gamma$, 
and for closure-based confidence boost $b>1$;
rule $X\to XY$ with $X\cap Y=\emptyset$, 
$c(X\to XY)\geq \gamma$, 
and $s(XY)\geq \tau$} 
\KwResult{boolean value indicating whether 
$\cl{\beta}(X\to XY) > b$}
mine $\D$ for the basis $\Bst{}$ at threshold $\gamma/b$\;
\For{each rule $X'\to X'Y'\in\Bst{\gamma/b}$ 
where $X'\cap Y'=\emptyset$, with
$X'\subseteq \cl{X}$ and $Y\subseteq \cl{X'Y'}$}{
\If{$\exists Z\subset \cl{X}-X'$ 
such that $\cl{X'Z}\subset\cl{X}$ (with inequality)
and $c(X\to XY)\leq b\times c(X'Z\to X'ZY)$}{
\Return{{\tt False}}}
\If{$\exists A \in X'Y'-\cl{XY}$ such that $c(X\to XY)\leq b\times c(X\to XAY)$}{
\Return{{\tt False}}}}
\Return{{\tt True}}
\caption{A variant of Algorithm~\ref{a:dconf} for closure-based 
confidence boost}\label{a:cbdconf}
\end{algorithm}

%
%
%
%
%
%
%
%
%
%
%
%
%
%
%
%
%
%

\begin{theorem}
\label{th:clalgo}
Let $X\to XY$ be a rule of confidence at least $\gamma$.
Algorithm~\ref{a:cbdconf} 
accepts it if and only if $\cl{\beta}(X\to XY) > b$.
\end{theorem}

\noindent
{\sl Proof.}
We follow essentially the same steps as in Theorem~\ref{th:algo},
although we must argue more carefully about the places
where the closure operator plays a role.
Again, we see first that the rejections are correct. 
In each case,
we just found
a rule $X''\to X''Y''$ with $X''\subseteq \cl{X}$ and $Y\subseteq Y''$,
be it $X'Z\to X'ZY$ or $X\to XAY$. 
In both cases, 
$(\cl{X}\neq\cl{X''} \lor \cl{XY}\neq\cl{X''Y''})$ holds:
in the first case, $\cl{X'Z} \neq \cl{X}$ is explicitly checked,
whereas, for the second case, $A\in XAY\subseteq\cl{XAY}$ but
$A\notin\cl{XY}$.
In each case, the rule $X''\to X''Y''$ 
contributes to
the
maximization in the denominator of the confidence boost and shows that
its value is less than or equal to~$b$. 

To see that acceptance is correct, 
assume $\cl{\beta}(X\to XY)\leq b$: 
we prove that, at some point, 
rule $X\to XY$ must fail one of the two tests
in the algorithm. By the definition of 
closure-based confidence boost, there must exist 
some rule $X''\to X''Y''$ 
with $X''\subseteq\cl{X}$, $Y\subseteq \cl{X''Y''}$,
and
$(\cl{X}\neq\cl{X''} \lor \cl{XY}\neq\cl{X''Y''})$,
and
such that $c(X\to XY)\leq b\times c(X''\to X''Y'')$.
Then,
from $c(X\to XY)\geq \gamma$ we infer $c(X''\to X''Y'')\geq \gamma/b$, so that
there must exist a rule in the basis $\Bst{\gamma/b}$, 
let it be $X'\to X'Y'$, that makes $X''\to X''Y''$ redundant
(possibly itself) under closure-based redundancy.
By Lemma~\ref{l:clredchar}, 
$X'\subseteq\cl{X''}$ and $X''Y''\subseteq\cl{X'Y'}=X'Y'$,
where the last equality is due to the fact that
$X'\to X'Y'\in\Bst{\gamma/b}$ so that $X'Y'$ is closed.
At some point (unless a correct
negative answer is found earlier), 
the algorithm will consider 
this rule $X'\to X'Y'\in\Bst{\gamma/b}$.
As in the proof of Theorem~\ref{th:char}, we distinguish two cases.

First assume that 
$\cl{X''}\subset \cl{X}$.
Since $X'\subseteq \cl{X''}$, we can consider 
$Z=\cl{X''}-X'\subset \cl{X}-X'$:
at some point, the algorithm will compare $c(X\to XY)$ to 
$b\times c(X'Z\to X'ZY)$. But it holds that $X'Z = \cl{X''}$ and that 
$Y\subseteq \cl{X''Y''}$, resulting in 
$c(X\to XY)\leq b\times c(X''\to X''Y'') = b\times c(\cl{X''}\to \cl{X''Y''}) 
\leq b\times c(X'Z\to X'ZY)$
and failing the test.

Alternatively, let's consider the case where 
$\cl{X''}\subseteq \cl{X}$ holds with equality: 
$\cl{X''} = \cl{X}$, so that $\cl{XY}\neq\cl{X''Y''}$; on the
other hand, we know now $X\subseteq\cl{X}=\cl{X''}\subseteq\cl{X''Y''}$, 
and also $Y\subseteq \cl{X''Y''}$, so that $\cl{XY}\subseteq\cl{X''Y''}$.

Assume briefly that $Y''\subseteq \cl{XY}$: 
as $X''\subseteq \cl{X''} = \cl{X}\subseteq\cl{XY}$,
we would obtain $\cl{X''Y''}\subseteq\cl{XY}$ and, therefore,
the equality $\cl{XY}=\cl{X''Y''}$; however, we know that 
this equality does not hold.

Hence, $Y''$ is not included in $\cl{XY}$, and there is some 
$A\in Y''\subseteq X'Y'$ that is not in~$\cl{XY}$,
that is, $A \in X'Y'-\cl{XY}$.
(If we know that $X=\cl{X}$, for instance when
the rule $X\to XY$ comes from a $\Bst{}$ basis,
$X'\subseteq\cl{X''} = \cl{X} = X$ tells us that
the search for $A$ can be circumscribed further to just
$A \in Y'-XY$.)
In due time, the algorithm will compare $c(X\to XY)$ 
to $b\times c(X\to XAY)$. But $\cl{X}=\cl{X''}$, and $A\in Y''$ so that 
$XAY\subseteq \cl{X''Y''}$, hence   
$c(X\to XY)\leq b\times c(X''\to X''Y'') = b\times c(\cl{X''}\to \cl{X''Y''}) 
\leq b\times c(X\to XAY)$
and the test will fail as well. This completes the proof.\qed

We report on a second algorithm below.


\subsection{Inequalities}

Compared to confidence boost, closure-based confidence boost 
relaxes the alternative rules to which a given rule is compared, 
e.g.~by allowing left hand sides included in $\cl{X}$ that are not
included in $X$; but, on the other hand, restricts them by the
proviso that the rules are ``inequivalent'' in a closure-based
sense, and not just different. Therefore, either can end up being
higher than the other, and the relationship with other quantities
like width or support ratio become less clear. 
We must review which inequalities still hold; we start
with the (partial) analogs of 
Propositions \ref{p:widthboost}~and~\ref{p:blift}.

\begin{proposition}
\label{p:suppclboost}
Assume $XY$ closed. Then,
the closure-based confidence boost 
is bounded by the support ratio:
$\cl{\beta}(X\to XY)\leq \sigma(X\to XY)$.
\end{proposition}

\noindent
{\sl Proof.}
Let $Z$ be the proper superset of $XY$ of largest support above $\tau$,
so that $\sigma(X\to XY) = s(XY)/s(Z)$. As $XY$ is closed, 
$\cl{Z}\neq\cl{XY}$. Rule $X\to Z$ enters, therefore, the
maximization in the denominator of the closure-based
confidence boost and leads to
$\cl{\beta}(X\to XY) \leq c(X\to XY)/c(X\to Z) = s(XY)/s(Z) = \sigma(X\to XY)$.\qed

\begin{proposition}
\label{p:cblift}
Assume
$s(X) < n$, the dataset size; then, the 
closure-based confidence boost 
$\beta(X\to XY)$ is bounded above by 
the lift of $X\to Y$.
\end{proposition}

\noindent
{\sl Proof.}
We consider the rule $\emptyset\to Y$.
For it to play a role in closure-based
confidence boost, we need $\cl{\emptyset}\neq\cl{X}$, 
which is equivalent to $s(X) < n$. The rest
of the argumentation is as in Proposition~\ref{p:blift}:
its support is above the threshold, and
$\cl{\beta}(X\to XY)\leq\frac{c(X\to XY)}{c(\emptyset\to Y)}$
which is the lift of $X\to Y$.\qed

It is interesting to note
that the condition about the left-hand side being nonempty
in Proposition~\ref{p:blift}
corresponds now to having support less
than the dataset size: the intuition is
that any items that appear in all transactions
become part of the closure of the empty set,
which is now the limit case.

We discuss now some relationships between the plain
and the closure-based versions of the confidence boost.

\begin{proposition}
\label{p:clboovsboorr}
Let $X\to XY$ be an association rule where $XY$
is a closed set and $X$ is a minimal generator.
Then, 
$\cl{\beta}(X\to XY) \leq \beta(X\to XY)$.
\end{proposition}

\noindent
{\sl Proof.}
Let $\beta(X\to XY)=b$: there must be a different rule 
$X'\to X'Y'$ such that $X'\subseteq X$, $Y\subseteq Y'$,
and $\frac{c(X\to XY)}{c(X'\to X'Y')} = b$. 
Assume first that $X'\subset X$.
As $X$ is a minimum generator, any subset
of $X$ has strictly larger support. Hence,
$s(\cl{X})=s(X)\neq s(X')=s(\cl{X'})$, which
implies that $\cl{X}\neq\cl{X'}$; then,
the same rule $X'\to X'Y'$ is accounted for 
in $\cl{\beta}$ as well, and leads
to a value of at most $b$. 

The remaining case is $X=X'$, which requires
that $XY\neq X'Y'$. Moreover, both $X=X'\subseteq X'Y'$
and $Y\subseteq X'Y'$ by the definition of confidence
boost, and $XY$ is closed, 
so that $\cl{XY} = XY\subset X'Y'\subseteq\cl{X'Y'}$.
Again in this case $X'\to X'Y'$ is accounted for 
in $\cl{\beta}$, and the stated inequality holds.\qed

\begin{corollary}
\label{c:ineqrr}
Let $X\to XY$ be a representative rule
at any confidence threshold; then
$\cl{\beta}(X\to XY) \leq \beta(X\to XY)$.
\end{corollary}

One interesting particular case is that of rules of
confidence~1 formed when
$X$ is a minimum generator of the closed set $XY$ itself; 
these rules form the
min-max exact basis 
from Definition~\ref{d:minmax} \cite{PasBas}
(a nonminimal basis for the implications of confidence~1, 
as the GD basis is sometimes smaller~\cite{GD}).
Proposition~\ref{p:clboovsboorr}
applies to these rules as well, of course.
On the other hand, we have:

\begin{proposition}
\label{p:clboovsboobst}
Let $X\to XY$ be an association rule where 
both $X$ and $XY$
are closed sets.
Then, 
$\beta(X\to XY) \leq \cl{\beta}(X\to XY)$.
\end{proposition}

\noindent
{\sl Proof.}
Let $\cl{\beta}(X\to XY) = b$: there must be a rule 
$X'\to X'Y'$ such that 
$\frac{c(X\to XY)}{c(X'\to X'Y')} = b$, fulfilling
the conditions 
$X'\subseteq \cl{X}$, 
$Y\subseteq \cl{X'Y'}$,
and either
$\cl{X}\neq\cl{X'}$ or 
$\cl{XY}\neq\cl{X'Y'}$.
We observe first that,
as $X$ is closed, 
$X'\subseteq \cl{X} = X$. 
Together with $X\cap Y=\emptyset$,
we get for later use that
$X'\cap Y=\emptyset$ as well.

We modify the rule 
$X'\to X'Y'$ 
by extending
its right-hand side into 
a closed set, as
$X'\to \cl{X'Y'}$,
which has the same confidence,
and then rewrite it into
$X'\to X'Y''$ 
by setting $Y'' = \cl{X'Y'} - X'$.
Note that $Y\subseteq \cl{X'Y'}$,
together with
$X'\cap Y=\emptyset$,
leads to $Y\subseteq Y''$.

Hence, with that rule written in this form,
the properties become
$\frac{c(X\to XY)}{c(X'\to X'Y'')} = b$, 
$X'\subseteq \cl{X} = X$, 
$Y\subseteq Y''$,
and either
$\cl{X}\neq\cl{X'}$ or 
$\cl{XY}\neq\cl{X'Y''}$.
It suffices to show that
$X'\to X'Y''$ and
$X\to XY$ are different rules
to ensure that $X'\to X'Y''$ 
participates in the computation 
of $\beta(X\to XY)$ and, hence,
to obtain the desired inequality.
But: if $\cl{X}\neq\cl{X'}$,
then necessarily $X\neq X'$; 
and, in the other case,
$XY = \cl{XY}\neq\cl{X'Y''} = X'Y''$
as both $XY$ and $X'Y'' = \cl{X'Y'}$
are closed sets. This completes the proof.\qed

As the $\Bst{}$ basis consists of rules where
both antecedent $X$ and consequent $XY$ are closed sets,
we obtain:

\begin{corollary}
\label{c:ineqbst}
Let $X\to XY$ be a rule in the $\Bst{}$ basis
(at confidence $c(X\to XY)$); then,
$\beta(X\to XY) \leq \cl{\beta}(X\to XY)$.
\end{corollary}

For the not unusual cases where a representative rule
participates 
as well
in the $\Bst{}$ basis, Section~\ref{s:cb}
suggests measuring its confidence boost, whereas
Section~\ref{s:cbcb} would propose to measure its
closure-based confidence boost. Now we see that
there is no conflict:

\begin{corollary}
\label{c:noconflict}
If $X\to XY$ is both a representative rule 
and a member of the $\Bst{}$ basis
(both at confidence $c(X\to XY)$), then
$\beta(X\to XY) = \cl{\beta}(X\to XY)$.
\end{corollary}

This follows at once from 
Corollaries \ref{c:ineqrr}~and~\ref{c:ineqbst}.

\begin{example}
In general, either of $\beta$ and $\cl{\beta}$ can
be strictly larger, when permitted by the statements
we have proved so far. 
In Example~\ref{ex:cbcb}, we saw a $\Bst{}$ rule
for which $\cl{\beta}(CDE\to AB) > \beta(CDE\to AB)$.
This also shows that Corollary~\ref{c:ineqrr} cannot be
extended to the $\Bst{}$ basis. Conversely,
as $\cl{A} = AB$ in our running example, rule
$B\to C$ is taken into account for the closure-based
confidence boost of the representative rule $A\to BC$,
leading to $\cl{\beta}(A\to BC) < 1$, whereas
$\beta(A\to BC) = 16/15$ as we saw in Example~\ref{ex:properineqs}.
\end{example}

We develop some further inequalities and yet another algorithm
that we will employ in Section~\ref{s:yacaree}.

\begin{theorem}
\label{th:usefulbound}
Assume that a threshold $b$ has been fixed for
the closure-based confidence boost.
Consider rule $X\to XY$ where both
$X$ and $XY$ are closed sets. Then 
$\cl{\beta}(X\to XY) \leq b$
if and only if either $\sigma(X\to XY) \leq b$,
or there is some closed proper subset $X'\subset X$,
$c(X\to XY) \leq b\times c(X'\to X'Y)$.
\end{theorem}

\noindent
{\sl Proof.}
Assume first $\cl{\beta}(X\to XY) \leq b$.
Let $X'\to X'Y'$ be the rule in the denominator
of the definition of $\cl{\beta}$ that leads to 
its actual value. 
Due to $Y\subseteq\cl{X'Y'}$, we have
$c(X'\to X'Y)\geq c(X'\to X'Y')$. 
If $\cl{X'}\neq\cl{X}$, as $X$
is assumed closed, we can state
$X'\subseteq\cl{X} = X$ so that, by monotonicity,
$X'\subseteq\cl{X'} \subset \cl{X} = X$.
Thus, $\frac{c(X\to XY)}{c(X'\to X'Y)} 
\leq \frac{c(X\to XY)}{c(X'\to X'Y')} 
= \cl{\beta}(X\to XY) \leq b$, 
and the second case holds. 
If, on the other hand,
$\cl{X'} = \cl{X}$, then $s(X) = s(\cl{X}) = s(\cl{X'}) = s(X')$
and, necessarily, $\cl{XY}\neq\cl{X'Y'}$;
yet $\cl{XY} = XY \subseteq X'Y' \subseteq\cl{X'Y'}$
as $XY$ is closed, hence $XY = \cl{XY}\subset\cl{X'Y'}$,
leading to 
$\sigma(X\to XY) \leq\frac{s(XY)}{s(X'Y')} =
\frac{c(X\to XY)}{c(X'\to X'Y')} = 
\cl{\beta}(X\to XY) \leq b$.

Conversely, if $\sigma(X\to XY) \leq b$
then $\cl{\beta}(X\to XY) \leq b$ by
Proposition~\ref{p:suppclboost}. Also, 
assuming $X'\subset X$ gives us
$c(X\to XY) \leq b\times c(X'\to X'Y)$,
where both $X$ and $X'$ are closed, 
$\cl{X'} = X'\subset X = \cl{X}$ 
so that $\cl{X'}\neq\cl{X}$, and 
rule $X'\to X'Y$ participates in
the computation of $\cl{\beta}(X\to XY)$,
leading to 
$\cl{\beta}(X\to XY) \leq 
\frac{c(X\to XY)}{c(X'\to X'Y)} \leq b$.\qed

For convenience in a later application,
we restate this theorem in its contrapositive form:

\begin{corollary}
\label{c:usefulbound}
Assume that a threshold $b$ has been fixed for
the closure-based confidence boost.
Consider rule $X\to XY$ where both
$X$ and $XY$ are closed sets. Then 
$\cl{\beta}(X\to XY) > b$
if and only if both $\sigma(X\to XY) > b$
and for every closed proper subset $X'\subset X$,
$c(X\to XY) > b\times c(X'\to X'Y)$.
\end{corollary}

Yet another application of this theorem is to
identify the analog of Proposition~\ref{p:liftboostsingle} 
for the closure-based case. To get there, it is
convenient to factor off the proof the following
technical but easy fact:

\begin{lemma}
\label{l:tech}
Let $X$ be a closed singleton, that is, $X=\cl{X}$ and $|X|=1$.
If $s(X) < n$, then there is exactly one closed proper subset
of $X$, namely $\emptyset = \cl{\emptyset}$; and, besides,
$X$ is free, that is, it is a minimum generator of itself.
\end{lemma}

\noindent
{\sl Proof.}
By definition, $\cl{\emptyset}$ contains exactly those items
that appear in all the transactions. By monotonicity, as
$\emptyset\subseteq Z$ for all $Z$, $\cl{\emptyset}$
is a subset of all closures. If $X$ is a closed singleton,
either $\cl{\emptyset} = \emptyset$ or 
$\cl{\emptyset} = X$; this second case is ruled out by
the condition $s(X) < n$, as $s(\cl{\emptyset}) = s(\emptyset) = n$.
Our statements follow.\qed

\begin{proposition}
Assume that 
$|X| = 1$ in rule 
$X\to XY$, that is, the left hand side is a single item.
Further, assume that $s(X) < n$, and that $X$ and $XY$ are closed.
Then $\cl{\beta}(X\to XY)$ coincides with the minimum among
the lift of $X\to Y$ and $\sigma(X\to XY)$.
\end{proposition}

\noindent
{\sl Proof.}
By Propositions \ref{p:suppclboost}~and~\ref{p:cblift},
we already know that $\cl{\beta}(X\to XY)$ is less than 
or equal to both quantities, under the given conditions.
To complete the proof,
we only need to show the
converse inequality, that is, 
$\cl{\beta}(X\to XY)$ is larger than or equal to
the minimum among
the lift of $X\to Y$ and $\sigma(X\to XY)$.
For this,
we will apply Theorem~\ref{th:usefulbound}:
$\cl{\beta}(X\to XY) \leq b$
if and only if either $\sigma(X\to XY) \leq b$
or there is some closed proper subset $X'\subset X$,
$c(X\to XY) \leq b\times c(X'\to X'Y)$.
We observe that, by Lemma~\ref{l:tech}, 
in our current conditions there is exactly one such $X'$, 
namely $\emptyset$, and the last inequality
becomes, then, the statement that the lift of $X\to Y$ is 
at most $b$; indeed, the lift coincides with
$\frac{c(X\to XY)}{c(\emptyset\to Y)}$.

As we can chose any value of $b$, we pick
simply $b = \cl{\beta}(X\to XY)$ itself, 
so that we can infer that either
$\sigma(X\to XY) \leq b = \cl{\beta}(X\to XY)$
or the lift of $X\to Y$ is also at most
$b = \cl{\beta}(X\to XY)$. Thus, either
$\sigma(X\to XY)$ or the lift of $X\to Y$
are less than or equal to $\cl{\beta}(X\to XY)$
and, certainly, the lesser of both quantities
obeys the same bound, which completes the proof.\qed


We obtain the corresponding variant of Corollary~\ref{c:quasitrackboost}:

\begin{corollary}
\label{c:trackboost}
Assume a threshold $b$ in place such that $\sigma(X\to XY)\geq b$ is
known, for $|X|=1$, that is, for a rule with a single antecedent item.
If $s(X) < n$, $X$ and $XY$ are closed,
and the lift of $X\to Y$ is less than $b$, then it equals
$\cl{\beta}(X\to XY)$. 
\end{corollary}

As a consequence, $\cl{\beta}(X\to XY) = \beta(X\to XY)$
for these cases. This is also consistent with 
Corollary~\ref{c:noconflict}: as we have stated in
Lemma~\ref{l:tech}, in this case $X$ is both closed
and a minimal generator; if $c(X\to XY) < 1$, then
this implies that it is equivalent to state that
$X\to XY$ is a representative rule and to state that
it is in the $\Bst{}$ basis.
This corollary will be very relevant in the implementation
described in Section~\ref{s:yacaree}.


\subsection{Alternative Algorithm}
\label{ss:altalg}

Theorem~\ref{th:usefulbound} leads to an alternative algorithm
to filter rules from the $\Bst{}$ basis according to their
closure-based confidence boost; we present it as
Algorithm~\ref{a:altcbcb}. Its correctness is 
immediate from Theorem~\ref{th:usefulbound}.
This algorithm 
is part of the tool described in Section~\ref{s:yacaree};
it tends to be better than the previous one when left-hand 
sides tend to be small. It pays the price of traversing 
all closed subsets of a given closed set but spares 
traversing the alternative basis at lower confidence.
In our implementation, as described below,
the test of the support ratio is actually pushed into
the closure mining, so that it becomes unnecessary 
to repeat it at the time of evaluating rules.

\begin{algorithm}
\DontPrintSemicolon
\KwData{dataset $\D$;
thresholds for support $\tau$, 
for confidence $\gamma$, 
and for closure-based confidence boost $b>1$;
rule $X\to XY$ with $X\cap Y=\emptyset$, 
$X$ and $XY$ both closed,
$c(X\to XY)\geq \gamma$, 
and $s(XY)\geq \tau$} 
\KwResult{boolean value indicating whether 
$\cl{\beta}(X\to XY) > b$}
\If{$\sigma(X\to XY) \leq b$}{\Return{{\tt False}}}
\If{$\exists Z\subset X$ closed such that
$c(X\to XY) \leq b\times c(Z\to ZY)$}{\Return{{\tt False}}}
\Return{{\tt True}}
\caption{An alternative algorithm for closure-based confidence boost}\label{a:altcbcb}
\end{algorithm}

\section{Empirical Validation}
\label{s:emval}

This section describes the outcomes of several empiric
applications of the notions of confidence boost;
the next section describes a complete tool that employs 
closure-based
confidence
boost, and the properties we have developed, to
offer parameter-less association mining.
With respect to specific datasets, 
we report first on objective figures: numbers 
of rules passing rather mild 
confidence boost thresholds on three datasets, all
consisting of real world data, but of very different 
characteristics. 
Subsequently, we briefly discuss the much more
difficult and subjective question of whether
the rules that we find are actually the rules
one may want.

\subsection{Quantitative Evaluation}
\label{ss:qeval}

\begin{table}
\tbl{Information about datasets}{
\begin{tabular}{|l|r|r|r|l|}
\hline
Dataset & Size  & Items & Occurrences \\
\hline\strut
\ds{Adult}
        & 32561 &   269 & 358171      \\
\ds{Retail}
        & 88162 & 16470 & 908576      \\
\ds{Now}
        &  1597 &  3873 &  14135      \\
\hline
\end{tabular}
}
\label{tb:datasets}
\end{table}

Dataset \ds{Adult} is the training set part of the Adult 
US census dataset from UCI~\cite{UCI}. 
Dataset \ds{Retail} was downloaded from the FIMI repository,
and contains typical market basket data
({\tt http://fimi.cs.helsinki.fi/}); and dataset \ds{Now}
(based on the Neogene of the Old World dataset, 
public release 030710~\cite{NOW}) 
is a transactional version of a paleontological dataset
from Europe: we downloaded and preprocessed 
slightly
file {\tt NOW\_public\_030710.xls}, 
so that
each paleontological site has been casted into a
transaction, where the items in the transactions are the species
of which fossile remains have been found at that site. Additional
information such as name or geographical position of the site have
been omitted, in order to keep the transactional format.

Table~\ref{tb:datasets} gives some information about the
datasets: their size (in number of transactions), the number of
items involved, and the total of item occurrences.
Each dataset has been mined at two different levels of support and
three different levels of confidence. Support thresholds were chosen
so as to produce noticeable numbers of rules, and also to make sure that 
the closure spaces were nontrivial in size (several thousand closures). 
Table~\ref{tb:numeritos} reports, for each pair of
support and confidence values, the basis size (RR/$\Bst{}$,
standing for representative rules and $\Bst{}$ basis
respectively)
and then the number of these basis rules, for each basis,
passing the corresponding confidence boost thresholds as given.
Of course, for the $\Bst{}$ case we bound the closure-based confidence boost.

\begin{table}
\tbl{Sizes of RR/$\Bst{}$ bases at confidence boosts 1 to 1.3}{
\begin{tabular}{|l||c|c|c||c|c|c||}
\hline
\ds{Retail}
 & $\tau$: 0.2\%  & $\tau$: 0.2\%  & $\tau$: 0.2\%  & $\tau$: 0.1\%  & $\tau$: 0.1\%  & $\tau$: 0.1\% \\
 & $\gamma$: 90\%  & $\gamma$: 80\%  & $\gamma$: 70\%  & $\gamma$: 90\%  & $\gamma$: 80\%  & $\gamma$: 70\% \\
\hline
basis &  111 / 111 &  205 / 205 &  572 / 572 &  248 / 233 &  652 / 643 &  1990 / 1984 \\
\hline
${}\geq1.00$ & 89 / 89 & 179 / 179 & 529 / 529 & 180 / 169 & 568 / 559 & 1819 / 1808 \\
${}\geq1.05$ & 26 / 26 & 87 / 87 & 384 / 384 & 44 / 43 & 327 / 323 & 1367 / 1362 \\
${}\geq1.10$ & 25 / 25 & 51 / 51 & 253 / 253 & 34 / 34 & 169 / 168 & 891 / 888 \\
${}\geq1.15$ & 25 / 25 & 35 / 35 & 150 / 150 & 34 / 33 & 113 / 112 & 545 / 543 \\
${}\geq1.20$ & 25 / 25 & 33 / 33 & 101 / 101 & 30 / 30 & 69 / 69 & 331 / 331 \\
${}\geq1.25$ & 24 / 24 & 32 / 32 & 63 / 63 & 27 / 27 & 53 / 53 & 178 / 178 \\
${}\geq1.30$ & 24 / 24 & 31 / 31 & 52 / 52 & 27 / 27 & 42 / 42 & 102 / 102 \\
\hline
\hline
\ds{Adult}
 & $\tau$: 5.0\%  & $\tau$: 5.0\%  & $\tau$: 5.0\%  & $\tau$: 2.5\%  & $\tau$: 2.5\%  & $\tau$: 2.5\% \\
 & $\gamma$: 90\%  & $\gamma$: 80\%  & $\gamma$: 70\%  & $\gamma$: 90\%  & $\gamma$: 80\%  & $\gamma$: 70\% \\
\hline
basis &  817 / 812 &  851 / 848 &  781 / 777 &  2288 / 2240 &  2090 / 2069 &  2004 / 1971 \\
\hline
${}\geq1.00$ & 308 / 290 & 281 / 274 & 316 / 309 & 898 / 823 & 729 / 698 & 803 / 768 \\
${}\geq1.05$ & 17 / 17 & 47 / 47 & 62 / 62 & 50 / 48 & 117 / 113 & 171 / 166 \\
${}\geq1.10$ & 7 / 7 & 17 / 17 & 33 / 33 & 19 / 18 & 51 / 50 & 82 / 81 \\
${}\geq1.15$ & 1 / 1 & 8 / 8 & 18 / 18 & 9 / 8 & 26 / 25 & 48 / 47 \\
${}\geq1.20$ & 0 / 0 & 3 / 3 & 12 / 12 & 4 / 3 & 14 / 13 & 35 / 34 \\
${}\geq1.25$ & 0 / 0 & 2 / 2 & 8 / 8 & 3 / 2 & 9 / 8 & 23 / 22 \\
${}\geq1.30$ & 0 / 0 & 2 / 2 & 8 / 8 & 1 / 0 & 7 / 6 & 19 / 18 \\
\hline
\hline
\ds{Now}
 & $\tau$: 0.4\%  & $\tau$: 0.4\%  & $\tau$: 0.4\%  & $\tau$: 0.3\%  & $\tau$: 0.3\%  & $\tau$: 0.3\% \\
 & $\gamma$: 90\%  & $\gamma$: 80\%  & $\gamma$: 70\%  & $\gamma$: 90\%  & $\gamma$: 80\%  & $\gamma$: 70\% \\
\hline
basis &  246 / 30 &  483 / 347 &  596 / 489 &  1646 / 30 &  2789 / 1710 &  3368 / 2443 \\
\hline
${}\geq1.00$ & 202 / 30 & 445 / 310 & 590 / 481 & 1302 / 30 & 2529 / 1295 & 3213 / 2189 \\
${}\geq1.05$ & 202 / 30 & 438 / 299 & 565 / 454 & 1302 / 30 & 2505 / 1255 & 3156 / 2104 \\
${}\geq1.10$ & 193 / 23 & 393 / 250 & 549 / 435 & 1284 / 23 & 2403 / 1126 & 3026 / 1971 \\
${}\geq1.15$ & 116 / 14 & 260 / 131 & 514 / 402 & 1097 / 14 & 2011 / 822 & 2602 / 1582 \\
${}\geq1.20$ & 108 / 14 & 242 / 120 & 466 / 359 & 526 / 14 & 1285 / 466 & 2049 / 1090 \\
${}\geq1.25$ & 91 / 9 & 204 / 94 & 431 / 327 & 500 / 9 & 1236 / 443 & 1991 / 1051 \\
${}\geq1.30$ & 76 / 4 & 184 / 85 & 404 / 308 & 473 / 4 & 1158 / 384 & 1842 / 929 \\
\hline
\end{tabular}
}
\label{tb:numeritos}
\end{table}

Our implementation was not particularly aimed at speed. Still,
for instance, computing all the figures regarding the representative
rule basis took less than 35 minutes on 
a low-range laptop. 
For the higher support threshold in each 
dataset, each computation time was between 20 and 45 seconds. 
For the larger, more demanding closure lattice at the lower 
support threshold of each dataset, these figures required 
between 2 minutes and
up to a maximum of 6 minutes. 
It will not be difficult to improve the running times 
in future work,
as a number of known accelerations
can be applied; we are already
undertaking this task. 
Computationally, the slowest part was always 
the construction of the closure lattice.

With respect to the outcome, we see that the reduction of the number 
of rules is clear, and in some cases it is very considerable. 
Recall that the bound at 1 of the confidence boost discards those 
basis rules for which a rule with {\em higher} confidence 
can be obtained by either reducing the antecedent, enlarging the 
consequent, or both; in the first case, it would mean that
the rule is actually a case of negative correlation that is
better left off from the output.

\subsection{Subjective Evaluation}
\label{ss:subjeval}

Quantitatively, the figures just given imply that large fractions of 
representative rules are somewhat uninteresting in that they 
fully lack any novelty, measured according to confidence boost. 
However, one may question
whether the 
actual 
rules passing the thresholds 
are ``the right ones''.
To our subjective
perception, after seeing the outcome of our experiments, the
whole process makes a lot of sense, but, in order to argue 
that indeed bounding the confidence boost leads to a
worthy data mining scheme, we should find a more 
convincing argumentation.
We hasten to add here that using the mined rules for
classification will not provide a reasonable evaluation, since for
such applications we must focus on single pairs of attribute and
value as right-hand side, thus making useless to consider larger
right-hand sides;
and, also, the classification
will only be sensible to minimal left-hand sides independently of
their confidences
(as in Subsection~\ref{ss:mmr} below).
Because of these properties, a classification task
is not fine enough to provide information about
the usefulness of the subtler confidence quotients 
involved in the confidence boost bounds. 

Clearly, the difficulty of this evaluation lies in the
fact that the issue is largely subjective.
At the present moment, our way through is to involve
``end-users'' in the evaluation of the obtained
association rules: persons that are
extremely well-versed on the dataset at hand.
Both for our version of confidence boost, and for a
sensible extension of it to handle absence of items
besides presence of items in the transactions, we are
developing an analysis of educational datasets, 
containing information about online courses on
multimedia systems and on the Linux operating system, 
in close cooperation with the teachers of said 
courses~\cite{BalTirZor10b}. Here, however, instead of 
looking for experts on a given dataset, we use a dataset 
for which some readers of this paper might be expected to be 
reasonably knowledgeable: 
in the same vein as the evaluations in \cite{MINI},
we employ the titles, topics, and abstracts of all 
the reports submitted to the {\em e-prints} repository 
of the Pascal Network of Excellence along its early years 
of existence. This dataset, extracted from the 
repository by Professor Steve Gunn, 
was the object of a visualization challenge 
of the Pascal Network in 2006. (Professor Gunn
has also kindly furnished to this author a
similar but much larger
dataset, to which we plan to apply the same 
scheme in the near future.)

\begin{table}
\tbl{Number of rules passing closure-based confidence boost bounds}{
\begin{tabular}{|l|r|r|r|r|r|r|r|r|r|r|r|}
\hline
Conf.& 1 & 1.05 & 1.1 & 1.15 & 1.2 & 1.25 & 1.3 & 1.35 & 1.4 & 1.45 & 1.5 \\
\hline
70\% & 948 & 824 & 689 & 554 & 417 & 331 & 247 & 175 & 142 & 112 & 85 \\
75\% & 639 & 541 & 444 & 356 & 266 & 212 & 161 & 112 & 97 & 76 & 56 \\
80\% & 367 & 298 & 231 & 182 & 132 & 101 & 78 & 54 & 43 & 36 & 26 \\
\hline
\end{tabular}
}
\label{tb:eprints}
\end{table}

The collection of papers was processed starting from a
plain text file containing one line for each of the 721
papers, including the title, the subjects chosen from
among the specific choices allowed by the repository
(marked by a '!' sign that we changed into the word
``subject''), and the whole text of the abstract of 
the report. The (mild) preprocessing consisted in 
removing punctuation and nonprintable characters, 
mapping all letters into lowercase, stripping off
stop words as per the list from {\tt www.textfixer.com},
and removing duplicate words from each of the 
transactions so obtained. This left 45185 total word
occurrences chosen from a vocabulary of 8233 items.
We checked the size of the closure space at supports
of 10\% (135 closures) and 5\% (830 closures, still
somewhat small), and then at 1\% (too large, as after a few 
minutes the program was still computing the closure
lattice's edges---in fact, a later run showed that 
it consists of 59713 closures). 
We settled for a far from trivial 
but manageable closure space consisting of 9621 closed 
itemsets obtained at 2\% support. Then, we computed 
the $\Bst{}$ basis at confidences 70\% (1070 rules),
75\% (729) rules, 
and 80\% (412 rules), and cut them down by filtering
them at closure-based confidence boosts of 1, 1.05, 1.1, 
1.15, 1.2, 1.25, 1.3, 1.35, 1.4, 1.45 and 1.5. 
All the runs were almost instantaneous.
The figures obtained, given in Table~\ref{tb:eprints}, make it
indeed possible to proceed to manual inspection of
many of these options.

\begin{table}
\tbl{Abbreviations of subjects for Tables 
\ref{tb:nuggets} and \ref{tb:gd} 
below}{
\begin{tabular}{|l|l|}
\hline
subject:BC & Brain-Computer Interfaces \\
subject:CI & Computational, Information-Theoretic Learning with Statistics \\
subject:IR & Information Retrieval and Textual Information Access \\
subject:LS & Learning/Statistics and Optimisation \\
subject:MV & Machine Vision \\
subject:TA & Theory and Algorithms \\
\hline
\end{tabular}
}
\label{tb:abbr}
\end{table}

Next, as a particular case, we chose to perform an examination 
of the 26 rules found at 2\% support, 80\% confidence, 
and 1.5 (closure-based) confidence boost, which revealed
rules with little or no redundancy among themselves, 
all of them semantically sensible, and with a handful of them
actually quite interesting (for this author). The whole
process leading to these ``nuggets'' lasted less than
two hours, {\em including all the preprocessing}, for
a single person (the author) and quite limited computing 
power (an old Centrino Solo laptop). These rules are
given in Table~\ref{tb:nuggets}. The predefined subjects
of the e-prints Pascal server appearing in the table
have been shortened to fit the page; Table~\ref{tb:abbr} 
reports the abbreviations used for them in 
Tables \ref{tb:nuggets}~and~\ref{tb:gd}. 

\begin{table}
\tbl{The 26 rules at 2\% support, 80\% confidence, 1.5 boost}{
\begin{tabular}{|r|r||l|c|l|}
\hline
conf. & supp \% & & & \\
\hline
 0.842 & 2.219 & principal & $\Rightarrow$ & component \\
 0.842 & 2.219 & unlabeled & $\Rightarrow$ & data \\
 0.882 & 2.080 & approach method show & $\Rightarrow$ & data \\
 0.850 & 2.358 & features selection & $\Rightarrow$ & feature \\
 0.842 & 2.219 & methods subject:MV & $\Rightarrow$ & images \\
 0.833 & 2.080 & nonlinear subject:LS & $\Rightarrow$ & learning \\
 0.810 & 2.358 & kernel used & $\Rightarrow$ & method \\
 0.889 & 3.329 & presents & $\Rightarrow$ & paper \\
 0.833 & 2.080 & solve & $\Rightarrow$ & problem \\
 0.941 & 2.219 & art & $\Rightarrow$ & state \\
 0.800 & 2.219 & brain & $\Rightarrow$ & subject:BC \\
 0.914 & 4.438 & document & $\Rightarrow$ & subject:IR \\
 0.907 & 5.409 & documents & $\Rightarrow$ & subject:IR \\
 0.826 & 2.635 & web & $\Rightarrow$ & subject:IR \\
 0.900 & 2.497 & feature learning & $\Rightarrow$ & subject:LS \\
 0.850 & 2.358 & features subject:TA & $\Rightarrow$ & subject:LS \\
 0.842 & 2.219 & linear problem & $\Rightarrow$ & subject:LS \\
 0.833 & 2.080 & data second & $\Rightarrow$ & subject:LS \\
 0.818 & 2.497 & data subject:MV & $\Rightarrow$ & subject:LS \\
 0.818 & 2.497 & more use & $\Rightarrow$ & subject:LS \\
 0.919 & 4.716 & object & $\Rightarrow$ & subject:MV \\
 0.895 & 4.716 & bound & $\Rightarrow$ & subject:TA \\
 0.889 & 5.548 & bounds & $\Rightarrow$ & subject:TA \\
 0.818 & 2.497 & graphs & $\Rightarrow$ & subject:TA \\
 0.813 & 3.606 & variables & $\Rightarrow$ & subject:TA \\
 0.813 & 10.264 & support & $\Rightarrow$ & vector \\
\hline
\end{tabular}
}
\label{tb:nuggets}
\end{table}

By way of comparison, at the same level of support, 
at the most demanding possible level of confidence
(100\%), with the less redundant basis computation 
currently known (the Guigues-Duquenne basis, \cite{GD}), the
result is 44 rules, with considerably more 
``intuitive redundancy''
and less interest overall, and requires somewhat 
longer time to be computed. 
Note that, by
their own definition, the rules in the $\Bst{}$ basis 
do not attempt at capturing rules with 100\% confidence,
but just at complementing them with partial rules;
hence, the Guigues-Duquenne basis has some additional
information. 
%
%
For the sake of comparison,
this basis is given in Table~\ref{tb:gd}. 
The considerable redundancy is clear:
many variants of ``support'' implies ``vector'' become
reduced to a single one under the confidence boost bound.
One may ask why the similar case of ``vector'' implies 
``support'' is missing from the list of 26 rules: the answer
is that its confidence is slightly under 75\% and, thus,
it is not reported under the 80\% threshold. Once more 
we see that setting the thresholds with no formal guidance
runs into very risky processes. It would be necessary 
to try and help the user by some sort of self-adjustment
of the thresholds. We have attempted at one first approach 
along this line, which is reported next.




\begin{table}
\tbl{The 44 implications in the Guigues-Duquenne basis at 2\% support}{
\begin{tabular}{|r||l|c|l|}
\hline
supp \% & & & \\
\hline
2.358 & al & $\Rightarrow$ & et \\
2.219 & machine models & $\Rightarrow$ & learning \\
2.219 & subject:LS support svms vector & $\Rightarrow$ & machines \\
2.358 & hidden markov & $\Rightarrow$ & models \\
2.080 & bci & $\Rightarrow$ & subject:BC \\
2.080 & eeg & $\Rightarrow$ & subject:BC \\
2.080 & collections & $\Rightarrow$ & subject:IR \\
2.219 & document paper & $\Rightarrow$ & subject:IR \\
2.358 & documents paper & $\Rightarrow$ & subject:IR \\
2.358 & document new & $\Rightarrow$ & subject:IR \\
2.497 & document documents & $\Rightarrow$ & subject:IR \\
2.774 & document information & $\Rightarrow$ & subject:IR \\
2.080 & data results vector & $\Rightarrow$ & subject:LS \\
2.497 & data learning problem set & $\Rightarrow$ & subject:LS \\
2.080 & object results & $\Rightarrow$ & subject:MV \\
2.219 & image images subject:LS & $\Rightarrow$ & subject:MV \\
2.358 & image object & $\Rightarrow$ & subject:MV \\
2.358 & images recognition & $\Rightarrow$ & subject:MV \\
2.358 & object recognition & $\Rightarrow$ & subject:MV \\
2.635 & images results & $\Rightarrow$ & subject:MV \\
2.774 & images object & $\Rightarrow$ & subject:MV \\
2.219 & algorithm generalization & $\Rightarrow$ & subject:TA \\
2.358 & bound subject:LS & $\Rightarrow$ & subject:TA \\
2.080 & based machines vector & $\Rightarrow$ & support \\
2.080 & machines used vector & $\Rightarrow$ & support \\
2.080 & paper show vector & $\Rightarrow$ & support \\
2.080 & classification machine vector & $\Rightarrow$ & support \\
2.080 & learning svm vector & $\Rightarrow$ & support \\
2.219 & machines svm vector & $\Rightarrow$ & support \\
2.358 & kernel machines vector & $\Rightarrow$ & support \\
2.497 & machines method vector & $\Rightarrow$ & support \\
2.497 & machines paper vector & $\Rightarrow$ & support \\
3.467 & machines using vector & $\Rightarrow$ & support \\
2.774 & machines such vector & $\Rightarrow$ & support \\
2.358 & machines svms & $\Rightarrow$ & support vector \\
2.080 & method problem support & $\Rightarrow$ & vector \\
2.080 & new subject:TA support & $\Rightarrow$ & vector \\
2.219 & support well & $\Rightarrow$ & vector \\
2.497 & machines methods subject:LS & $\Rightarrow$ & vector \\
2.635 & support svms & $\Rightarrow$ & vector \\
2.635 & learning machines subject:LS & $\Rightarrow$ & vector \\
2.913 & machines subject:LS subject:TA & $\Rightarrow$ & vector \\
3.606 & kernel support & $\Rightarrow$ & vector \\
6.380 & machines support & $\Rightarrow$ & vector \\
\hline
\end{tabular}
}
\label{tb:gd}
\end{table}

\section{Towards Parameter-Free Association Mining}
\label{s:yacaree}

In this section we describe an open-source software tool
that profits from closure-based confidence boost and its
properties to offer a sensible association mining process,
while refraining from asking the user to select any value of
any parameter:
our system {\sl yacaree} (Yet Another Closure-based Association
Rule Experimentation Environment), 
a proof-of-concept currently implemented fully in pure Python.
It combines several processes using lazy evaluation 
by means of the functional programming facilities available in 
current versions of Python to mine high-boost $\Bst{}$ 
association rules. Its key property is the self-tuning
of the support and the confidence boost thresholds.

As in most current proposals, {\sl yacaree} mines only frequent 
closed itemsets; initially, it enforces a support bound that 
starts ridiculously low (namely, at 5 transactions). In most 
applications, one cannot rely on mining all frequent closures 
at this threshold: this might or might not be possible, depending 
on the dataset; therefore, along the process, the threshold
will be automatically increased. Frequent closures are mined 
via a simplified variant of ChARM \cite{ZakiHsiao}, 
rather close to a depth-first search but with the proviso that 
closed itemsets are produced in order of decreasing support,
so that increasing the support threshold does not invalidate 
the closures found so far.

This idea is reminiscent of the decreasing support in the 
version of ``apriori'' implemented in the Weka tool \cite{Weka}, 
but in that well-known system the user still has to provide 
a maximum and a minimum values to try the support threshold,
and a ``delta''  by which the support keeps decreasing; then, the 
``apriori'' algorithm is run repeatedly for the corresponding 
sequence of support thresholds. Further, the process stops when 
a given number of rules, also chosen by the user, has been found. 
This makes it unlikely to find rules of low support. The 
``predictive apriori'' alternative, present in that tool as well 
\cite{Scheffer,Weka}, also attempts at adjusting the support, 
by balancing it with respect to confidence. Our system 
works very differently, as it is able to mine closures in order 
of decreasing support by its own algorithmics, and self-adjusts 
the internal effective support bound on the basis of technological 
limitations, in a manner that is autonomous and independent of the 
confidence or of any other parameter of the mining process.

The closed set miner takes the form 
of an iterator, and searches for the next closed set to be 
reported only when asked to do so. Each closure found is 
analyzed, upon yielding it to the next phase, to see 
whether it can be further extended without failing the
current support threshold, and all those extensions,
with their explicit supporting transaction lists, are added 
to a heap which provides instantaneously the largest-support 
closed set that has not been extended so far.

The closures are passed on to a lattice constructor, a 
``border'' algorithm which computes the lattice structure, 
so that immediate predecessors of each closed set are 
readily available, as it is convenient for computing 
the basis $\Bst{}$. 
The lattice constructor itself is based on \cite{BSVG} 
and works also as an iterator, constructing Hasse edges 
only when they are needed. Rules are, then, constructed 
from the lattice. Closures and candidate rules are either 
discarded, if we can guarantee that future threshold adjustments 
will never recover them; or processed, if they obey the 
thresholds; or maintained separately on hold, if they fail the current 
thresholds but might turn to obey them after future adjustments. 


The support threshold changes along the process.
It starts, as indicated, at an almost trivial level, 
and grows, if necessary, as the monitorization of the 
mining process reveals that the 
memory consumption surpasses internal thresholds. 
More precisely, 
the heap where unexpanded closures are stored
is considered in overflow when either its length, 
or the total memory it uses, or the sum of the lengths of 
the associated support lists, exceeds a corresponding 
predefined threshold. At that point, the minimal support 
constraint is recomputed and raised as necessary so that 
the exploration can continue. In this way, both the risk
of entering a huge closure space, and the risk of memory
overflow upon computing the supports of the closed sets
(as sometimes happens for dense datasets) are avoided.

We impose a very mild confidence threshold that
remains fixed, letting large quantities of rules pass;
but we control the number of rules to be provided to the 
user via a threshold on the closure-based confidence boost,
which is adjusted also along the run.
We use the
approximation to the confidence boost provided by the support
ratio (Proposition~\ref{p:suppclboost})
to push the confidence boost constraint into 
the mining process,
and we use the lift, applied to the particular cases 
to which 
Corollary~\ref{c:trackboost}
applies,
to self-adjust the boost threshold. 

In fact, as the Hasse edges of the closures lattice are identified,
the support ratio can be
computed easily. If it is lower than the current confidence
boost threshold, the closure is not adequate to yield high boost rules,
but it could become so if, in the future, the 
confidence boost
threshold
decreases. Therefore, 
the confidence boost constraint is partially
``pushed into'' the mining process by
temporarily omitting the expansion of such closed sets. 
Instead, they are maintained 
separately into a dedicated data structure, from where they are 
``fished off'' again in case a decrease of the boost bound
promotes them to candidate closures for creating high-boost rules.
We take advantage of the support ratio constraint also to compute
the confidence boost of rules, as per Algorithm~\ref{a:altcbcb}:
we know that, if the closed set reaches that stage, then its
support ratio is high enough, so we do not need to test it again.



The mining process starts with a somewhat
demanding confidence boost bound, that requires a rule to have 
at least 15\% more confidence than any of 
the rules participating in its confidence boost
in order to qualify as interesting. In some datasets, this 
figure is not that restrictive, and dozens of rules still make it.
By default, the system writes off as 
result the up to 50 rules of highest boost.

In many datasets, though, that confidence boost bound is too
demanding. The program monitors the 
lift of 
rules having one single item as antecedent and 
obtained
from a closed set that has support ratio above the confidence
boost bound
(cf.~Corollary~\ref{c:trackboost}).
If these 
lift values keep decreasing, they
enter a weighted average with the current 
confidence boost bound and 
may
decrease
it. 
In this way, we track
the degree of correlation empirically found 
in the dataset to reduce conveniently the confidence boost bound.
There is a static limit to this boost bound: it is never allowed
to drop below 1.05.
(All the hardwired limits can be modified easily 
in the same module {\tt statics.py} of the source code.)

The result is a functional preliminary system, where ample room 
still remains for efficiency and algorithmic improvements, which
shows that it is possible to find interesting association
rules in a fully autonomous manner: the user simply selects
a dataset and launches the process, which takes just one to
five minutes in many easy datasets, and up to ten to twenty 
minutes on a modern laptop for a few difficult, highly dense 
datasets. The output is a set of rules which,
in most cases, is reasonably small and 
shows independent and sensible associations.

\begin{figure*}
\includegraphics[width=\textwidth]{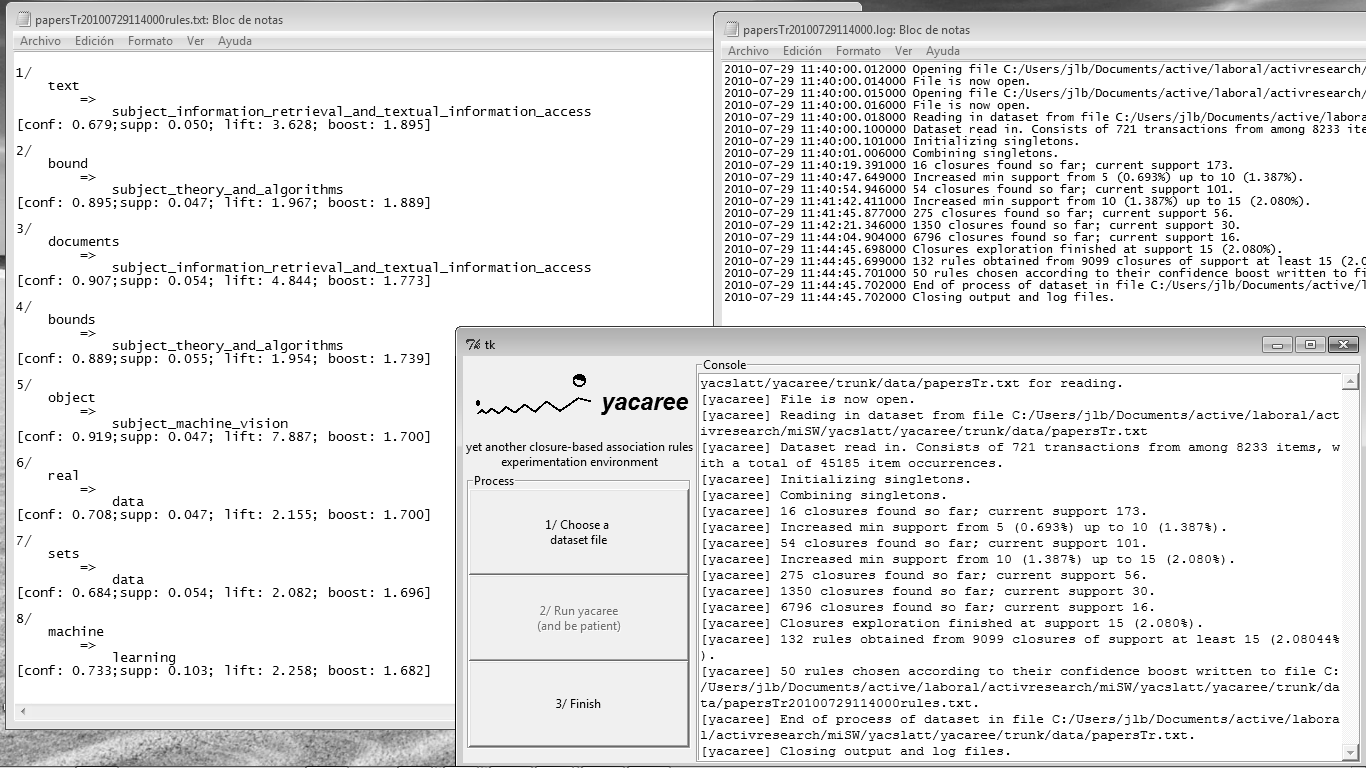}
\caption{A screenshot of {\sl yacaree} with the rules and log output files}
\label{fg:scr}
\end{figure*}

The open source, plus some example datasets, 
can be downloaded from 
{\tt http://sourceforge.net/projects/yacaree/};
these example datasets 
are already preprocessed 
into transactional form, 
and come from \cite{UCI} or
\cite{NOW}, or from the e-prints repository of the Pascal
Network of Excellence. The screenshot provided in 
Figure~\ref{fg:scr} shows the simple
interface (button ``Run'' is disabled as the system has
been just run) and the two text files generated: the log,
where we can see that the process took a bit over five
minutes, and the start of the file containing the rules found. Both
the console and the log indicate the self-adjustments
of the support; 
along this particular run,
no adjustment was performed on the boost threshold,
as enough high-boost rules were found for its initial value.


\section{Discussion} 
\label{s:discussion}

The main contribution of this paper is the
closure-based confidence boost: a new concept
that measures a form of objective novelty for
association rules, which we have studied from 
the formal and algorithmic perspective and which
we have used to construct open source association
mining tools.

Our starting point was the study of notions of
redundancy in a ``logical'' spirit. When a rule
is irredundant, we still can use relative 
confidences to assess the degree of irredundancy,
which we see as a potentially useful formalization
of objective novelty.

A redundancy due to larger consequents can be
measured by the support ratio; as such, both earlier
notions like confidence width and our new proposals
are related to it. A redundancy due to smaller 
antecedents only in some cases is handled appopriately 
by the preexisting confidence width, due to the 
stringent condition of ``logical'' redundancy; 
with the also preexisting notion of blocking, 
the case of smaller antecedents is handled in a 
less strict, more intuitively useful way. A bound
on the simplest of the two versions of confidence 
boost is exactly equivalent to bounding both 
preexisting notions, width and blocking; therefore,
our first new proposal allows for much smoother
handling of the combination of the previously
studied concepts.


As the notion of plain confidence boost 
turns out to be debatable for
one specific ``closure-aware'' basis, the $\Bst{}$ rules,
we have proposed also a more sophisticate ``closure-aware''
version of the confidence boost, for which we have
developed the corresponding formal and algorithmic
study. 

An obvious drawback of using a confidence boost bound is the need
to choose yet another parameter for the mining process, besides
confidence and support. However, in our experiments, this problem
did not seem to be that serious:
a noticeable aspect of the confidence boost bound is that
the outcome of the mining shows relatively quite low
sensitivity both with respect to its precise value and 
with respect to the values of other parameters such as
confidence: quite similar sets of rules are obtained. 
We quickly learned to use two standard values,
at 1.05 to prune off just really low novelty rules and at 1.2 to prune
more aggressively; whereas, in case the dataset still gives
many rules above this threshold, occassionally we would employ the very
drastic value of 1.5. This scheme tends to work well, and not only
that: it also
make less critical the choice of the confidence threshold, that
can be safely left at a somewhat low value (say, around 0.6 to 0.7),
leaving to the boost parameter the task of reducing the output size.
These empirical facts were widespread to such an extent
that we attempted at using (closure-based) confidence boost
to try and construct a parameter-free association miner:
the {\sl yacaree} system, able to self-tune the closure-based
confidence boost and the support thresholds.
We believe that
the embodiment 
of the computation of the~$\Bst{}$ basis 
together with closure-based
confidence boost bounds
in an open source tool
will promote 
its use in data mining practice, as {\sl yacaree}
exhibits a unique quality
of ``turnkey'' system that works with just the few clicks needed
to choose the input dataset. Of course, it can be used as well
in the standard manner, as the default initial values of 
confidence, support, and other internal parameters can be 
manually tuned effortlessly, if necessary, by data mining 
experts. However, this action is not anymore necessary, as
{\sl yacaree} is ready to do its best with no need of user 
choices.
The system is platform independent, although in a system 
with small memory, the control of the heap size may require 
some initial tuning (to be made just once) to avoid runtime 
errors for lack of memory; whereas, in very powerful systems, 
obtaining the most of them may also require some tuning. 

The shortcomings of confidence thresholds discussed at the beginning
of Subsection~\ref{ss:block} have been often interpreted as an
inadequacy of the very notion of confidence. Yet, we prefer to develop 
our proposal in the context of support and confidence bounds, for several 
reasons.

First, conditional probability is a concept known to many educated 
users from a number of scientific and engineering disciplines, so that 
communication between the data mining expert and the domain expert 
is often simplified if our measure
is confidence. Second, as a very elementary concept, it is the best playground 
to study other proposals, such as our contribution here, which could be then 
lifted to other similar parameters. 

Third, and more importantly, we believe that, in fact, our approach of
complementing it with relative measures will make up for many of the
objections raised against confidence. In fact, our interpretation of this
sort of objections 
is not the widespread consequence that ``confidence is 
inapproprate'' to filter and rank association rules, but that 
``an \emph{absolute}
threshold on confidence is inappropriate'' to filter and rank association
rules. This does not mean that it has to be replaced as a measure of
intensity of implication, and, in fact,
it has been observed and argued that 
(at least in somewhat sparse transactional datasets)
the combination of support and
confidence is already very good at discarding rules that are present only 
as statistical artifacts and do not really correspond to correlations in the 
phenomenon at the origin of the dataset \cite{MeggSrik}; instead, we 
consider that our message is that it should be complemented with 
\emph{relative}
confidence thresholds that assess the novelty of each rule by comparison
with the confidence of logically (or intuitively) stronger rules. The
identification of the precise notion for this task is a clear research
issue, to which we have contributed via our two variants of the 
notion of confidence boost.

A number of connected approaches 
to association rule quality 
exist in the literature.
We discuss here those that we have found most closely
related; Subsection~\ref{ss:mmr}
is devoted to the 
deeper analysis of a particularly close
contribution. 
We finish the paper with a description of forthcoming work.

\subsection{Comparisons to Related Work}
\label{ss:relwork}

We refer to \cite{GH} for an excellent survey of many
options to relate supports of left and right hand sides
of association rules
to construct indicators of interestingness. Many of these
only work on a single rule, with no reference to alternative
rules with, say, smaller but otherwise arbitrary left-hand
sides. A notable case is lift, which implicitly refers to
a rule with the same right-hand side and an empty left-hand
side, as discussed in the proof of Proposition~\ref{p:blift}.
Compared to this family of measures, confidence boost is
finer as it can distinguish among many alternative antecedents
to compare, at the price of being potentially more expensive
to evaluate due to the search for smaller but arbitrary 
left-hand sides, and larger but arbitrary right-hand sides.
We have shown several algorithms that attempt at circumscribing
this search to smaller spaces.

More sophisticated interestingness measures are possible,
for instance those based on the KL-divergence between 
probability distributions induced with and without the 
given rule~\cite{JarSim}: the induced distributions 
satisfy the supports of the rule and of its antecedent 
but otherwise maximize the entropy. In preliminary tests,
our approach, with quite robust settings of confidence
(between 0.6 and 0.7) and boost (stardard threshold of~1.2)
gives results very close to those in~\cite{JarSim}.

Several published works attempt at a similar detection of the
``exceptionality'' or ``surprisingness'' of rules; many of these
work in the relational setting, instead of the transactional
setting where our work fits. Relational data can be analysed
in the transactional setting by converting a pair given by an
attribute name and a value for the attribute into a single
item, as we do in the \ds{Adult} dataset in Table~\ref{tb:numeritos}.
Assuming the relational structure of the data, however, brings in
the extra power of ``implicit negation'' of attributes, due to 
the incompatibility among simultaneous values of the same
attribute. This implicit negation is useful to explain novelty 
by comparing more specific rules stating a consequent of the form 
$A=V$ to more general rules stating a consequent of the form $A=V'$ 
for $V'\neq V$, and quite interesting results along this line can be 
found in \cite{PadTu2000,Suz,SuzKod}, among others.
Our purely transactional setting (like for the \ds{Retail} 
or \ds{Now}
datasets) does not allow us to employ this method of implicit negation 
and, therefore, such contributions are not directly comparable to ours.

A few additional contributions that still lie in the
transactional setting and are similar to ours are discussed next.
The notions of confidence width and rule blocking from 
\cite{Bal09} are similar to the ``pruning'' proposal from \cite{LiuHsuMa}, 
in that the intuition is the same; also our proposal here follows an 
analogous intuitive path. Major differences are that, in the proposals 
we discuss, a large portion of the pruning becomes unnecessary because 
we work on minimum-size bases, namely representative rules, 
and, more importantly, 
that the pruning in~\cite{LiuHsuMa} is based on the $\chi^2$ statistic,
whereas we will look instead into the confidence thresholds that would 
make the rule ``redundant'', either in a ``formal logic'' sense or in 
a more intuitive, but still logical-style relaxation.
Our notions are also similar to the notion 
of {\em improvement}, proposed in \cite{BAG} and also discussed 
in \cite{LiuHsuMa,Webb07}; 
but improvement is a measure of an absolute, additive 
confidence increase, with no reference to representative rules or redundancy, 
and it only allows for varying the antecedent into a smaller one, keeping the 
same consequent. 

\subsection{Minimum Antecedent and Maximum Consequent}
\label{ss:mmr}

Many works suggest further notions of redundancy, in most cases
based upon mere intuition. The fact that a rule $X\to XY$
is redundant with respect to $X\to XY'$ whenever $Y\subset Y'$
(in the sense of having at least the same confidence)
is pointed out in many places 
(e.g.~\cite{AgYu,KryszPAKDD,PhanLuongICDM,ShahLaksRS}). 
Our starting point
being the representative basis, we only would keep $X\to XY$
if its confidence is higher than that of $X\to XY'$, by a 
factor indicated by the confidence boost; this quantification
is an effective refinement of that known proposal. 

On the other hand, 
redundancy 
of $X\to XY$ with respect to $X'\to X'Y$, where $X'\subset X$, is 
debatable.
As we have already discussed in Subsection~\ref{ss:block},
rules $X\to XY$ and $X'\to X'Y$, where $X'\subset X$, 
provide different, orthogonal information.
Still, one may wish to forget about $AB\to C$ if $A\to C$ is 
already present; this seems a natural attitude, and, in fact, 
explicit proposals of removing the seemingly redundant
rule appear in many references, often jointly with the (correct)
observation of redundancy due to larger consequents.
This happens in the structural cover of \cite{ToKleRHM}, and in 
some of the pruning rules of \cite{ShahLaksRS} (which focuses 
on a slightly different approach since their main measure 
is actually lift, but, in fact, most of their developments work 
for confidence as well); and also in \cite{Scheffer}.
All these proposals may make sense as heuristics, and their
connection to confidence boost is developed below; however,
if taken as redundancy statements then they are incorrect 
and, in some cases, where a 
precise mathematical statement and its proof are provided 
(like~\cite{Scheffer}),
the proof can be seen to switch into a ``full implication'' 
meaning of the ``arrow'' connective, and is actually wrong, therefore, 
since it does not apply to partial rules.
Discarding the apparently weaker rule requires more care
and a finer discussion and, actually, the confidence boost 
provides for this.

In fact, without pretending to argue redundancy, one could 
consider rules with minimal antecedent and maximal consequent 
simply as an heuristic for handling a large set of mined
rules, acting as a sort of summaries
of rules with larger antecedents or shorter consequents, 
or both. As a representative of these 
proposals, we chose to discuss the approach of~\cite{KryszPKDD} 
which can be casted as follows:

\begin{definition}
\label{def:mmr}
For a fixed confidence threshold $\gamma$ 
and a fixed support threshold~$\tau$,
the \emph{minimal-antecedent, maximal-consequent rules} MMR${}_{\tau,\gamma}$
are those rules $X\to XY$ (with $X\cap Y = \emptyset$) such
that $c(X\to XY)\geq\gamma$, $s(X\to XY)\geq\tau$, and for which
the following holds: the only rule $X'\to X'Y'$ with
$X'\cap Y' = \emptyset$, $c(X'\to X'Y')\geq\gamma$, 
$s(X\to XY)\geq\tau$ which satisfies that 
$X'\subseteq X$ and $Y\subseteq Y'$, is itself: 
$X=X'$ and $Y=Y'$.
\end{definition}

The following holds \cite{KryszPKDD}:

\begin{proposition}
\label{p:mmrrr}
For a confidence threshold $\gamma$ and a support threshold $\tau$,
all MMR${}_{\tau,\gamma}$ rules are representative rules 
for these thresholds.
\end{proposition}

Let us point out that these rules are subtly different
from the min-max approximate basis of \cite{PasBas},
given in Definition~\ref{d:minmax}, their apparent
similarity notwithstanding.
There, the closed set forming the whole right-hand side 
is to be maximal, including the antecedent; here, only
the part of the closed set that does not belong to the
antecedent is to be maximal. As the antecedent is itself
minimal, the notions differ. In a sense, MMR are to
min-max rules as confidence boost is to confidence width.

\begin{example}
In our running example, we find that rule $BC\to A$
has confidence $\gamma=8/9$. It is a representative rule 
at its confidence threshold $\gamma=8/9$, hence it
is a min-max rule by Proposition~\ref{p:rrclos};
but it is not in MMR${}_{\tau,\gamma}$ since 
$c(B\to A) = 10/11 > \gamma$. This example also 
proves that the converse of Proposition~\ref{p:mmrrr} 
does not hold.
\end{example}

As discussed in depth in Subsection~\ref{ss:block},
we must be aware that MMR's may lose information, since rules 
that have nonminimal antecedents may be actually irredundant 
and potentially interesting. Our main proposal in this paper,
confidence boost,
can be interpreted as a quantitative variant of MMR's,
whereby nonminimal antecedents or nonmaximal consequents
are likely to be considered not novel (and conversely),
yet this connection depends on how well the rule clears
the confidence and support thresholds. More precisely:

\begin{proposition}
Fix support and confidence thresholds $\tau$ and $\gamma$.
\begin{enumerate}
\item
If $X \to Y$ is a MMR${}_{\tau,\gamma}$ rule, then
$\beta(X\to Y) \geq 
\min\left(\frac{s(X\to Y)}{\tau},\frac{c(X\to Y)}{\gamma}\right)$.
\item
If $X \to Y$ is \emph{not} a MMR${}_{\tau,\gamma}$ rule, then
$\beta(X\to Y) \leq \frac{c(X\to Y)}{\gamma}$.
\end{enumerate}
\end{proposition}

\noindent
{\sl Proof.}
\begin{enumerate}
\item
Consider an MMR${}_{\tau,\gamma}$ rule $X \to Y$. Any different
rule $X'\to Y'$ with $X'\subseteq X$ and $Y \subseteq Y'$ must
fail either the support threshold~$\tau$ or the confidence 
threshold~$\gamma$.~First we show that, for such a rule, 
$c(X'\to Y')\leq\max(\frac{\tau}{s(X)},\gamma)$,
considering two cases. Assume $X'\neq X$, 
and consider rule $X'\to Y$, which is also
different from $X\to Y$. We have 
$s(X'Y) \geq s(XY) > \tau$ so that it must fail
the confidence threshold; hence, 
$c(X'\to Y') \leq c(X'\to Y) < \gamma \leq \max(\frac{\tau}{s(X)},\gamma)$.
Assume now $X'=X$: either $c(X'\to Y') < \gamma$, 
or $X'\to Y'$ fails the support threshold, 
$s(X'Y') = s(XY') \leq \tau$, whence 
$c(X'\to Y') = \frac{s(X'Y')}{s(X')} \leq \frac{\tau}{s(X')} = \frac{\tau}{s(X)}$;
thus $c(X'\to Y')\leq\max(\frac{\tau}{s(X)},\gamma)$ again.

Now we can bound the confidence boost easily:
any rule considered for the maximization in the
denominator of the definition of confidence boost
has confidence at most $\max(\frac{\tau}{s(X)},\gamma)$,
and there are finitely many of them,
so that the denominator itself obeys the same bound,
which implies that 
$\beta(X\to Y) \geq 
\min\left(\frac{c(X\to Y)}{\frac{\tau}{s(X)}},\frac{c(X\to Y)}{\gamma}\right) =
\min\left(\frac{s(X\to Y)}{\tau},\frac{c(X\to Y)}{\gamma}\right)$.
\item
This part is quite simple. 
If $X \to Y$ is \emph{not} an MMR${}_{\tau,\gamma}$
rule, then
there must exist some different
rule $X'\to Y'$ with $X'\subseteq X$ and $Y \subseteq Y'$ 
passing the support and confidence thresholds; this rule
enters the maximization in the denominator of the
definition of confidence boost, which is, then, at least $\gamma$,
resulting in 
a confidence boost $\beta(X\to Y) \leq \frac{c(X\to Y)}{\gamma}$.\qed
\end{enumerate}

That is: a rule that is not an MMR${}_{\tau,\gamma}$ rule, and barely
clears the confidence threshold~$\gamma$, can be appropriately pruned
as not novel due to low boost; but, if its confidence is much
higher than the threshold, even if it is not MMR, it 
may exhibit
enough novelty to make it debatable whether it must be pruned off
the output. Conversely, an MMR${}_{\tau,\gamma}$ rule that clears 
barely the support and confidence thresholds may turn out
to be of low confidence boost, and it could be better to
omit it from the output. Essentially, the same purpose is
attempted by both approaches but confidence boost bounds
offers a quantitative evaluation of the extent to which
representative rules are appropriate as rules to choose for
the output of the mining process: they will often coincide
with the MMR${}_{\tau,\gamma}$ but these will be occassionally 
inadequate.


\subsection{Further Work}

Of course, the use of confidence boost does not preclude a combination 
with lift or any other measure of intensity of implication; to what 
extent these separate measures interact with confidence boost, and 
which ones perform best, is one among many open lines of future research.

Indeed, whatever method is proposed to reduce the output
of an association miner leaves a major doubt:
are these the rules one really wants?
We plan to continue 
working on this 
rather subjective
issue, and intend to employ 
further actual end-user evaluations 
from dataset providers, as we have started to do
with respect to partial aspects.
We are working on datasets coming from an e-learning platform, for 
which we have a manually recorded labeling of the interest of each rule, 
provided by the dataset suppliers, 
namely, the teachers of the courses where the datasets originated, who
are also available for consultation.
The particular characteristics of
this dataset require us first to extend our approach into handling
both presence and absence of each item
\cite{BalTirZor10b,BalTirZor10a}.
Also, sometimes, some 
of the full-confidence
implications would be desirable indeed for inclusion in the output, given
that working on the basis $\Bst{}$ leaves them fully out;
however,
it is unclear whether confidence boost would still be the right
notion, and, even so, full-confidence implications require to
compute the minimal generators of each closure, therefore losing
the desirable advantage offered by closure-based confidence boost
operating on top of $\Bst{}$ rules, which can be computed much
faster since they only use the closures lattice. We continue to 
investigate this problem, and some partial progress, on which
we still hope to improve, is reported in \cite{BalTirZor10a}.

The {\sl yacaree} tool 
has 
many developments open to further work. First, 
since we mine frequent closures in descending support, instead
of ascending, some of the optimizations in ChARM require further
work before being readily applicable; also, the best algorithm
in \cite{BSVG} (namely iPred) to compute Hasse edges is not applied, 
as it assumes a cardinality-ordered traversal of the closed sets 
instead of a support-oriented one; the theorems that guarantee 
its applicability have been obtained only recently, and
a forthcoming version of {\sl yacaree} will sport this faster
algorithm, iPred.
Also, it seems 
possible that a smarter coupling of the miner with the lattice
computation might provide further accelerations. 
On the other hand, from the point of view of the user,
and beyond efficiency improvement considerations,
a few
alternative internal configurations 
of the parameters might reveal
themselves useful, provided one can hit with intuitive 
descriptions that make them clearly understandable by 
nonexperts: indeed, whereas the user is grateful for
being able to run the program with no parameter selection,
{\sl yacaree} is not snake~oil, and it is likely that, for
certain datesets, and after seeing the result, the user may be 
tempted to ``try again'' in some alternative way. 



Hence, we will work next on improving the
speed of the system,
on finding sensible ways of reporting
interesting full-confidence implications
without paying too much as a time overhead,
and on developing interactions with 
end users to study their evaluations of the generated 
sets of rules, possibly leading thus to further refinements 
of the confidence boost notion and of any other aspect
that might be considered.
In the meantime, researchers interested in conducting 
their own evaluation can download the system freely and
analyze the output of confidence-boost-bounded 
mining on their datasets; this author would be grateful
to be informed of the results.


 
\bibliographystyle{acmsmall}

\bibliography{bibfile}

\end{document}